\documentclass[a4paper]{cas-sc}

\usepackage[authoryear,longnamesfirst]{natbib}
\usepackage{float} % for floats and the H placement marker
\usepackage{algorithm} % for Algorithms
\usepackage{algpseudocode} % for psedo code inside Algorithms
\usepackage{physics} % for \vb ("vector bold" characters)
\usepackage{scalerel,amssymb} % for the \msquare command
\usepackage{xcolor} % for coloring math formulas
\usepackage{caption} % for correctly formatting table captions
\usepackage{multirow} % for double-row tables
\usepackage{flafter} % prevent floats from floating up the page to before the section they were supposed to be in
\usepackage{url}

\def\tsc#1{\csdef{#1}{\textsc{\lowercase{#1}}\xspace}}
\tsc{WGM}
\tsc{QE}

\makeatletter
\renewcommand*\env@matrix[1][\arraystretch]{%
  \edef\arraystretch{#1}%
  \hskip -\arraycolsep
  \let\@ifnextchar\new@ifnextchar
  \array{*\c@MaxMatrixCols c}}
\makeatother

\floatname{algorithm}{Algorithm}

\NewDocumentCommand{\highlight}{O{blue!40} m m}{%
\draw[mycolor=#1] (#2.north west)rectangle (#3.south east);
}

\NewDocumentCommand{\fhighlight}{O{blue!40} m m}{%
\draw[myfillcolor=#1] (#2.north west)rectangle (#3.south east);
}

\makeatletter
\newenvironment{breakablealgorithm}
  {% \begin{breakablealgorithm}
     \refstepcounter{algorithm}% New algorithm
     \hrule height.8pt depth0pt \kern2pt% \@fs@pre for \@fs@ruled
     \renewcommand{\caption}[2][\relax]{% Make a new \caption
       {\raggedright\textbf{\fname@algorithm~\thealgorithm} ##2\par}%
       \ifx\relax##1\relax % #1 is \relax
         \addcontentsline{loa}{algorithm}{\protect\numberline{\thealgorithm}##2}%
       \else % #1 is not \relax
         \addcontentsline{loa}{algorithm}{\protect\numberline{\thealgorithm}##1}%
       \fi
       \kern2pt\hrule\kern2pt
     }
  }{% \end{breakablealgorithm}
     \kern2pt\hrule\relax% \@fs@post for \@fs@ruled
  }
\makeatother

\newcommand{\pluseq}{\mathrel{+}=}

\newdefinition{definition}{Definition}
\newtheorem{theorem}{Theorem}
\newtheorem{lemma}{Lemma}
\newdefinition{rmk}{Remark}
\newproof{pf}{Proof}
\newdefinition{ex}{Example}

\begin{document}
\let\WriteBookmarks\relax
\def\floatpagepagefraction{1}
\def\textpagefraction{.001}

\shorttitle{Greedy vine building}    

\shortauthors{D. Pfeifer and E. A. Kovács}  

\title [mode = title]{Trunc-Opt vine building algorithms}

\tnotemark[tnote mark] 

%\tnotetext[tnote number]{tnote text} 


\author[1]{Dániel Pfeifer}[type=editor]

%\cormark[1]


\ead{pfeiferd@math.bme.hu}

\credit{Writer, Algorithm implementation, Algorithm optimization}

%%%%%%%%%%%%%%%%%%%% FIEK?????????????????????

\affiliation[1]{organization={Budapest University of Technology and Economics},
            addressline={Műegyetem rkp. 3},
            city={Budapest},
%          citysep={}, % Uncomment if no comma needed between city and postcode
            postcode={1111}, 
            state={},
            country={Hungary}}

\author[2]{Edith Alice Kovács}

% Footnote of the second author

% Email id of the second author
\ead{kovacsea@math.bme.hu}


\credit{Writer, Conceptualization, proofs, algorithm}

\affiliation[2]{organization={Department of Differential Equations, Budapest University of Technology and Economics},
            addressline={Műegyetem rkp. 3}, 
            city={Budapest},
%          citysep={}, % Uncomment if no comma needed between city and postcode
            postcode={1111}, 
            state={},
            country={Hungary}}

%\cortext[1]{Corresponding author}

%\fntext[1]{footnote text}

\nonumnote{}

\begin{abstract}
Vine copula models have become highly popular and practical tools for modelling multivariate probability distributions due to their flexibility in modelling different kinds of dependences between the random variables involved. However, their flexibility comes with the drawback of a high-dimensional parameter space. To tackle this problem, truncated vine copulas were introduced by \cite{kurowicka2010optimal} (Gaussian case) and \cite{brechmann2013risk} (general case). Truncated vine copulas contain conditionally independent pair copulas after the truncation level.
So far, in the general case, truncated vine constructing algorithms started from the lowest tree in order to encode the largest dependences in the lower trees. The novelty of this paper starts from the observation that a truncated vine is determined by the first tree after the truncation level (see \cite{kovacs_szantai}). This paper introduces a new score for fitting truncated vines to given data, called the Weight of the truncated vine. Then we propose a completely new methodology for constructing truncated vines. We prove theorems which motivate this new approach. While earlier algorithms did not use conditional independences, we give algorithms for constructing and encoding truncated vines which do exploit them. Finally, we illustrate the algorithms on real datasets and compare the results with well-known methods included in R packages. Our method generally compare favorably to previously known methods.
\end{abstract}


% Research highlights (max 85 chars incl spaces)
\begin{highlights}
\item Proving that the first tree after truncation of a vine copula determines the truncated vine copula distribution
\item Proving that a regular cherry tree distribution can be expressed as a truncated vine distribution
\item Introducing a new score for characterizing the goodness of fit of truncated vine copula - weight of truncated vine copula
\item New greedy vine building algorithm that exploits conditional independences with the aim of minimizing KL-divergence - equivalent to maximizing the weight of truncated vine copula
\item Algorithm for encoding cherry tree sequences into a vine matrix usable by the R VineCopula package
\end{highlights}

% Keywords

\maketitle

% Main text
\section{Introduction}\label{section:intro}
% kiindulás, cél, dolgozat főcélja / felépítése, melyik részben mi van (max. egy oldal)

%In many applications, simple D-vines have been used so far to read pair-copula densities (p.d.f.'s) from. (see Figure \ref{d_vine}) \cite{aas_dvine} \cite{kurowicka_dvine} 
Vine copulas are widely used models, applied in hydrology (see \cite{paprotny2025evolution}), machine learning (see \cite{you2025escaping}), anomaly detection in time series (see \cite{chunquan2025ct}), finance (see \cite{nagler2019model}), and so on. Vine copulas allow the separate modeling of marginal probability distributions and dependences between the components of the multivariate probability distribution. Additionally, they can accommodate tail asymmetries and various forms of multivariate dependence simultaneously. This is achieved by constructing multivariate copulas using only bivariate (conditional) copulas in their expressions.
While the catalogue of bivariate parametric copula families is large, this is not the case for larger dimensions. The motivation for vine copula models was to construct multivariate copulas using only bivariate copulas as building blocks. \cite{joe1996families} gave the first pair copula construction in terms of distribution functions, while \cite{bedford1} developed constructions in terms of densities. They also provided a framework representing all possible constructions graphically in \cite{bedford2}. 
A significant advancement was that \cite{vinematrix} created a compact representation of the vine structure as a triangular matrix.

The flexibility of vine copula models comes with huge computational challenges. To tackle this problem, special vine structures were used, like C-vines and D-vines, which were used in many applications, see for example \cite{aas_dvine} and \cite{kurowicka_dvine}. Recently an interesting result was published by \cite{joe2025random}, where a method for generating random correlation matrices based on partial correlation C-vines was introduced.
In earlier studies examining a more general vine structure, two key observations were made. First, it is advantageous to model strong correlations in the initial trees. Second, in the subsequent trees, independence copulas tend to emerge. This observation eventually led to a greedy vine building method, in which trees are built greedily using the Kendall coefficient in \cite{Dissmann}. The method also relies on  Morales-Napoles's matrix representation introduced in \cite{vinematrix}. 

An alternative methodology for vine building was proposed by Kurowicka specifically for the case of the Gaussian copula in \cite{kurowicka2010optimal}, called sometimes the bottom-down method. This method utilizes partial correlations in its approach and takes advantage of conditional independences.
D. Kurowicka was the first to introduce the term "truncated vine", in \cite{kurowicka2010optimal}, which refers to a vine structure with independent copulas in all trees from a given level $t$ and higher. Although elliptical copulas are more flexible in the sense of pairwise dependence modelling, large correlation matrices are cumbersome to specify and may lead to over-parameterized models. However, truncated vines allow for a drastic reduction of the number of parameters. In addition, for elliptical copulas, the dependence in the joint lower and joint upper tails is the same. Truncated copulas were later also discussed by \cite{brechmann2013risk}, but even earlier they appear in Eike Brechmann's Diploma thesis \cite{brechmann2010truncated}. This important work utilizes the observation that in higher-level trees, conditional pair-copulas often become independence copulas. At the core of Brechmann's algorithm stands Dissmann's greedy algorithm, to retain of the strongest dependences in the lower trees. \cite{brechmann2013risk} proposed a sequential approach to the truncation challenge. They iteratively construct R-vine copulas, which are 1-truncated, 2-truncated, and so on. In each step, they use the Vuong test for model comparison to assess the gain of extending the truncated R-vine copula by an additional tree. If the gain is determined to be negligible, the procedure is stopped and the R-vine copula is truncated at the current level. A disadvantage of this method is that in each step, a truncated R-vine copula has to be fully specified, that is, a vine tree has to be determined, pair copulas have to be selected, and parameters have to be estimated. Due to the related computational effort, this hinders an efficient exploration of the search space of truncated R-vine copula models, and the authors may settle for a locally optimal result in each step of their procedure.

\cite{hanea2015non} introduced another method that also applies conditional independences in Bayesian networks. Another recent study by \cite{gauss2025properties} shows that in the case of no truncation, and if the parameter space increases with the number of variables $n$, consistency can still be shown.

In May 2011, in München the 4th Workshop on Vine Copula Distributions and Applications took place, where E. Kovács and T. Szántai drew the attention of the audience to the fact that the truncated vine copula is in fact cherry tree copula, which encodes the conditional independences, and it is important to set the truncation level as low as possible. Their idea was to fit a cherry tree probability distribution to the data, which defines the structure of the highest tree. They published their work first on arXiv \cite{kovacs2011vine}, then in a conference paper \cite{cherry1}. They showed that the trees in a vines do in fact have cherry-tree structures. This observation was also used in later written papers like \cite{brechmann2014parsimonious}, where these structures were renamed as Markov trees. This paper aimed to find low truncation levels based on the exploitation of conditional independences. In their paper, they discussed the special case of Gaussian distributions, where they used partial correlations in the construction of the trees of the vines. They proved that if data comes from a multivariate normally distributed random vector with zero mean vector and unit variances (possibly after standardisation), then the optimal (with respect to maximum likelihood estimation) $\ell$-truncated R-vine copula with vine tree structures $T_1,...,T_{\ell}$  and Gaussian pair copulas is obtained by minimising
$
    \sum_{i=1}^{\ell} \sum_{e \in E_i} \log \left( 1 - r_{j(e), k(e); D(e)}^2 \right)
$. For the corresponding notation see \cite{brechmann2014parsimonious}, formula (3.3). This result stands only for Gaussian probability distributions. Even in this case, for finding the best fitting $\ell$-truncated vine, there do not exist any exact optimization algorithms. Therefore, they introduced some heuristics to start not from the best first tree; instead, they started with more spanning trees, in the so-called 1-neighbourhood of the maximum/minimum spanning tree. 
Based on these results, \cite{brechmann2015truncation} developed a construction methodology by proposing a sequential approach that iteratively considers $\ell$-truncated models with $\ell= 1,2,...$ to find the smallest truncation level that meets a desired fit index, which was calculated based on a Gaussian distribution. The new truncation rule exploits the optimality results of the former paper \cite{brechmann2014parsimonious}, on the truncation of vine copulas with Gaussian building blocks. The goodness of fit of such vine copulas is measured by so-called fit indices, as used in the literature on structural equation modeling. Using such fit indices, they truncate such that a certain closeness to the best possible (nontruncated) case is achieved. They also obtain bounds on what is ignored through truncation. The newer approach separates the selection of the vine trees and the selection of bivariate copulas as building blocks. The selection of the truncation level is computationally very efficient and enables them to better explore the search space of truncated vine copulas, while the existing approaches before relied on locally optimal greedy algorithms. They propose to use a selection procedure inspired by genetic-kind algorithms to select flexible vine copulas with yet small truncation levels. 

Another significant but less well-known contribution is by \cite{kovacs_szantai}. They showed that in the case of continuous cherry tree probability distributions (a notion related to probabilistic graphical models), there exists a cherry tree copula density associated to the same cherry tree structure which defines the density. Although expressions from both of the methods contain conditional independences between the variables, the truncated R-vines constructed greedily based on \cite{brechmann2013risk}'s algorithm do not exploit the conditional independences existing in the data. The new idea introduced by \cite{kovacs_szantai} for modeling general truncated R-vines was to firstly construct, the first tree after truncation by exploiting conditional independences through fitting a cherry-tree copula to the data. After this, they gave an algorithm to fill out the lower trees of the vine. They did an optimization only for finding a cherry tree copula that fits the data. While their approach is more general since it is not restricted to the multivariate normal distribution, its downside is that discretization is  required because the optimization problem could only be solved based on it.

The present paper aims to give an algorithm for constructing truncated vines by exploiting conditional independences existing in the data, which is an advantage over Brechmann's truncated vine algorithm given in \cite{brechmann2013risk}. The algorithm we introduce here can be applied not solely for multivariate normal probability distribution or only Gaussian copulas, like the algorithms introduced in \cite{kurowicka2010optimal}, \cite{brechmann2014parsimonious}, \cite{brechmann2015truncation}. Our new algorithm does not require discretizations unlike in \cite{kovacs_szantai}, since an approximation of the information content will be introduced to the case of multivariate continuous random variables.
Also an important feature of the algorithm is that the structure can be encoded in a matrix, which makes it compatible with the very popular R-vine copula libraries. 
The goodness of fit to the data of the obtained truncated R-vines will be characterized by two methods: based on their log-likelihood and Kullback-Leibler divergence. In high-dimensional sparse vine copula models, other measures were introduced, see \cite{nagler2019model}.

The paper consists of 7 sections. The second section presents preliminaries such as information theoretical concepts, relations which will be used, the cherry tree graph structure, the cherry tree probability distribution, vine copulas and their representations. The third section contains important results on how truncated vine copulas relate to cherry tree copulas. The fourth section proves how a cherry tree copula can be obtained from a truncated vine copula. The fifth section presents the new TruncOpt algorithm, along with the introduction of an approximation for the information content based on data, and calculates the complexity of the TruncOpt methodology. In the sixth section, Brechman's greedy algorithm is compared to the new TruncOpt algorithm on different datasets by calculating the log-likelihood of the structure and the Weight of the truncated vine. Finally, the seventh section summarizes the paper.

\section{Preliminaries}\label{section:prelim}
In this section, we present notations and earlier results we will use throughout the paper. The first subsection presents the information theoretical concepts, the second one introduces cherry tree probability distributions, finally, the third subsection introduces vine copulas, their representations and how they are and related to cherry tree graphs.

\subsection{Information theoretical concepts and results}\label{subsection:info}
% Edith
% információtartalom
% cherry fa minimalizálja a Kullback-Leibler divergenciát

%Let us start from a sample dataset, $\boldsymbol{D} \in \mathbb{R}^{m \times n}$, $\boldsymbol{D} = [x_{i,j}]_{i \in \{1,\dots,m\},j \in \{1,\dots,n\}}$, with $m$ rows and $n$ columns, where each $n$-long row $\textbf{x} \in \boldsymbol{D}$ represents a realization from the probability distribution of the random vector $\textbf{X} = (X_1,\dots,X_n)$, such that each row is independent from one another and identically distributed.
We consider random vector $\textbf{X} = (X_1,\dots,X_n)$, the \textbf{joint cumulative distribution function} (cdf) of $\textbf{X}$ will be denoted by $F(\textbf{x})$, and the corresponding \textbf{joint probability density function} (pdf) by $f(\textbf{x})$.

The \textbf{entropy} of the joint probability distribution of $\textbf{X} = (X_1,\dots,X_n)$ is given by the following formula:

$$H(\textbf{X}) = - \int_{\mathbb{R}^n} f(\vb{x}) \log_2(f(\vb{x})) d\vb{x}.$$
It is important to mention here that the entropy could be negative for continuous random variables.

Then the \textbf{information content} of $\textbf{X} = (X_1,\dots,X_n)$ is defined as the difference between the sum of the marginal entropies and the total entropy:

\begin{align}
\label{infcontent}
I(\textbf{X}) = \left(\sum_{i=1}^n H(X_i)\right) - H(\textbf{X}) 
\end{align}

%In the continuous case this can be written as
%\begin{equation}
%    I(\textbf{X})= -\left(\sum_{i=1}^n
%\int_{\mathbb{R}} f_i(x_i) \log_2(f_i(x_i)) dx_i\right) + %\int_{\mathbb{R}^n} f(\vb{x}) \log_2(f(\vb{x})) d\vb{x}
%\end{equation}
If $X_1,\dots,X_n$ are independent then $\sum_{i=1}^n H(X_i) = H(\textbf{X})$, so $I(X)=0$. Due to the Jensen inequality, for any $\textbf{X}$, $I(\textbf{X}) \ge 0$ (see Theorem 2.6.3 in \cite{cover1999elements}). Therefore, $I(\textbf{X})$ is a quantification of dependence. For two random variables, it coincides with the concept of mutual information. (Information content was also called as multiinformation or multicorrelation by other authors.)

It is worth also noting that information content is a special case of the Kullback-Leibler divergence between a given joint probability distribution and the probability distribution defined by the product of the univariate marginals, which is the independent distribution since they have the same marginals as the initial joint probability distribution.

\subsection{Cherry tree graph structure and cherry tree probability distribution}\label{section:cherrydefs}
Cherry tree graph structures for modeling discrete probability distributions, by exploiting conditional independences, were first introduced in \cite{kovacs_szantai2} and then generalized in \cite{szantai2012hypergraphs}. Later, they introduced cherry tree copula \cite{cherry1} and \cite{kovacs_szantai}. Here we recall the results that we will use in the present paper.

We call a graph structure obtained by the following two steps a \textbf{cherry tree graph structure of order \textbf{k}}:
\begin{enumerate}  
\label{$k$-th order cherry tree}
    \item The smallest cherry tree is a clique graph of size $k$.  
    \item A cherry tree on $m+1$ vertices is obtained from a cherry tree on $m$ ($m>k$) vertices by connecting a new vertex to $k-1$ vertices that form a clique inside the cherry tree.
\end{enumerate} 

\begin{center}
\begin{figure}[ht]
    \centering
    \includegraphics[width=0.9\textwidth]{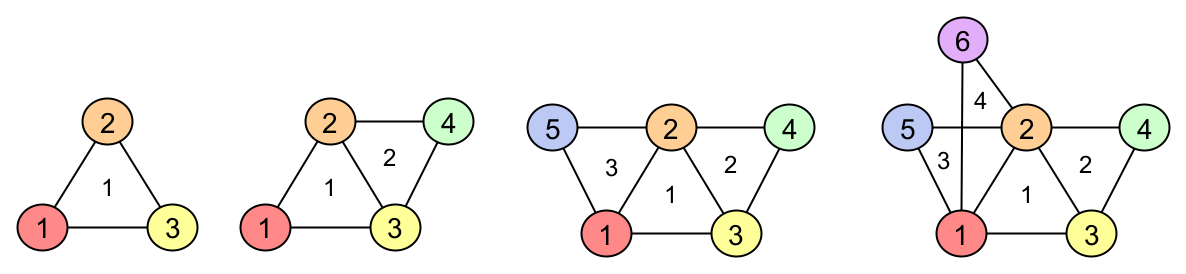}
    \caption{An example construction of a third-order cherry tree.}
    \label{fig:cherry_example}
\end{figure}
\end{center}

Figure \ref{fig:cherry_example} shows how a third-order cherry tree is built on 6 vertices step by step.

It is easy to see that a cherry tree is a triangulated (chordal) graph with the maximum cliques containing $k$ elements. Based on this, we introduce the concept of the cherry-junction tree. Intuitively, this can be seen as a cluster tree, where the clusters consist of $k$ elements, and the edges comprise $k-1$ elements that are common to the connected clusters.

\begin{figure}[ht]
    \centering
    \includegraphics[width=0.8\textwidth]{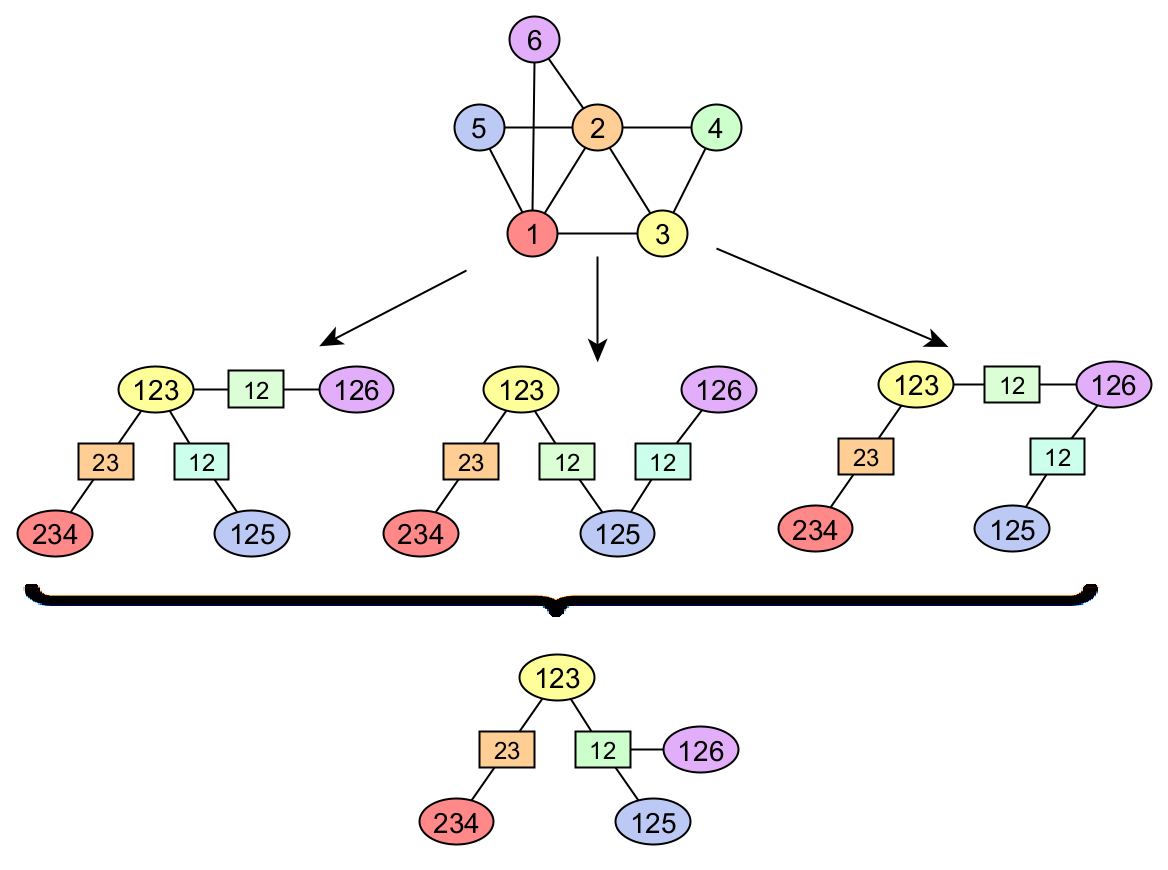}
    \caption{A regular cherry tree structure represented as a chordal graph (first row), junction trees (second row) and represented in a compact junction tree form (third row)}
    \label{fig:cherry_junction}
\end{figure}

We call the structure assigned to a cherry tree graph of order $k$ a \textbf{cherry-junction tree of order \textbf{k}}, and define it as follows:
\begin{enumerate}
\item  The set of vertices in each $k$th order maximum clique is called a \textbf{cluster}, and contains its $k$ vertices. 
\item Two $k$-element clusters are \textbf{connected} if the following two conditions are fulfilled:
\begin{itemize}
\item the clusters share $k-1$ elements
\item if a set of $k-1$ elements is contained by $m$ clusters, these clusters will be connected tree-like by $m-1$ edges, see second row of Figure \ref{fig:cherry_junction}.
\end{itemize}
\item The $k-1$-element set given by the intersection of two connected clusters is called a \textbf{separator}. The number of clusters containing it is the \textbf{multiplicity of the separator}.
\end{enumerate}
A cherry-junction tree is characterized by the set of indices $V$, the set of clusters $\mathcal{K}$, the set of separators  $\mathcal{S}$ and a dictionary of the multiplicities of the separators $\mathcal{M}$.

In the present paper, we will use only a subset of cherry trees which fulfil the following special property:

We call cherry-junction trees with the property that each of their clusters is connected to other clusters by at most two separators \textbf{regular cherry trees}.

For an illustration of a 3rd order regular cherry tree, corresponding to a 3rd order cherry tree graph, see Figure \ref{fig:cherry_junction}. The figure indicates that the junction tree associated to a cherry tree graph structure does not necessarily have a unique graph representation.

An important property of the junction trees is the \textbf{running intersection property}, i.e. if a vertex is contained in two different clusters it is contained in any cluster on the path between the two clusters. For a good overview on the relations between these concepts, see \cite{pfeifer2024vine}.

\vspace{2mm}
Let $\mathbf{X}=\left( X_{1},\ldots ,X_{n}\right)^{T}$ \ be a random vector
and  $V$ $=\{1,...,n\}$ be the set of indices, and let us consider a cherry-junction tree over $V$ given by the set of clusters $\mathcal{K}$ and the set of separators $\mathcal{S}$ denoted by $(V,\mathcal{K},\mathcal{S},\mathcal{M})$. 
For any marginal probability distribution $( X_{i_1},\ldots ,X_{i_k})$, we use the following abbreviation: $f(\mathbf{X}_{i_1,\ldots ,i_k}) = f(\mathbf{X}_A)$ where $A=\{i_1,\ldots ,i_k\}$.

\begin{definition}\label{def:cherry_tree_density}
   The\textbf{ cherry tree density function} of \textbf{X}, assigned to a cherry-junction tree $(V,\mathcal{K},\mathcal{S},\mathcal{M})$ is given by a joint density function $f_{ch}(\mathbf{x})$ by the following formula:
\begin{equation}
\label{eq:Cherry_tree}
f_{ch}( \mathbf{x}) =\frac{\prod\limits_{K\in \mathcal{K}}f_K\left( \mathbf{x}_{K}\right) }{\prod\limits_{S\in \mathcal{S}}f_S\left( \mathbf{x}_{S}\right)
^{v_{S}-1}}
\end{equation}
where $v_{S}\in\mathcal{M}$ is the multiplicity of $S$.
\end{definition}

A central theorem with respect to cherry tree approximations, regarding the KL-divergence between the real distribution $f(\mathbf{x})$ and the approximation assigned to a cherry tree denoted by $f_{ch}(\mathbf{x})$ is given by the following formula of \cite{kovacs_szantai}:

\begin{equation}
\label{KL_folyt}
\text{KL}(f_{\text{ch}},f)=\sum_{i=1}^n H\left( X_{i}\right) - H\left( 
\textbf{X}\right) - \left(\sum_{K\in \mathcal{K}} I(\textbf{X}_{K}) - \sum_{S\in \mathcal{S}} (\nu_{s}-1) I(\textbf{X})\right)
\end{equation}

We highlight here that only the expression in the parentheses of Formula \ref{KL_folyt} depends on the cherry tree structure, therefore, for a good fit to the data in order to minimize KL divergence, we have to maximize the expression in the parantheses. 

\subsection{Vine copulas and their representations}
\label{subsection:vine_copulas}
We start this subsection by recalling some concepts and theorems of Copula theory. Then we recall the definition of vine copulas and the representations needed for this paper.
Copulas were introduced by \cite{sklar}. 

%Since $F_i$ is the c.d.f. of $X_i$, the distribution of $F_i(X_i)$ is uniform on $[0,1]$. This can easily be seen from:

%$$\mathbb{P}(F_i(X_i) \le x_i) = \mathbb{P}(X_i \le F^{-1}(x_i)) = F(F^{-1}(x_i)) = x_i$$
\begin{definition}
\item A function $C: [0,1]^n \to [0,1]$ is an $n$-dimensional copula function, if it satisfies the following properties:

\begin{itemize}
\item $C(u_1,\dots,u_n)$ is strictly increasing in all $u_i$ components.
\item $C(u_1,\dots,u_{i-1},0,u_{i+1},\dots,u_n) = 0$ for all $u_k \in [0,1]$, $k \ne i$, $i \in \{1,\dots,n\}$.
\item $C(1,\dots,1,u_i,1,\dots,1) = u_i$ for all $u_i \in [0,1]$, $i \in \{1,\dots,n\}$.
\item $C$ is $n$-decreasing, meaning that for all $(u_{1,1},\dots,u_{1,n})$ and $(u_{2,1},\dots,u_{2,n})$ in $[0,1]^n$, if for all $i$, $u_{1,i} < u_{2,i}$, then

$$\sum_{i_1 = 1}^2 \cdots \sum_{i_n = 1}^2 (-1)^{\sum_{j=1}^n i_j} C(u_{i_1,1},\dots,u_{i_n,n}) \ge 0$$
\end{itemize}
\end{definition}

%Let $(U_1,\dots,U_n) = (F_1(X_1),\dots,F_n(X_n))$, where all the $U_1,\dots,U_n$'s have $\text{UNI}[0,1]$ distribution separately. Then the $C$ copula function can be defined the following way:

%$$C(u_1,\dots,u_n) = \mathbb{P}(U_1 \le u_1,\dots,U_n \le u_n)$$

%which is the joint c.d.f. of $(U_1,\dots,U_n)$.

%In the case where $X_1,\dots,X_n$ were independent, the joint distribution of the $U_i$'s will simply be a multivariate uniform distribution. However in the case where $X_1,\dots,X_n$ are not independent, we could get a different distribution. Because of this, $C(u_1,\dots,u_n)$ "expresses the dependence" between $(X_1,\dots,X_n)$.

Sklar's theorem \cite{sklar} can be regarded as a central theorem in copula theory as it relates joint cumulative probability distribution functions to copula functions:
\begin{theorem}
\label{sklar_theorem} Any multivariate c.d.f. $F$ can be written in the following way:

\begin{equation}\label{eq:sklar}
F(x_1,\dots,x_n) = C(F_1(x_1),\dots,F_n(x_n))
\end{equation}

where $C$ is a copula function. Moreover, if the one-dimensional marginals $F_1,\dots,F_n$ are continuous, then $C$ is unique.
\end{theorem}

Using Sklar's theorem, the $C$ copula c.d.f. can be calculated the following way:

$$C(u_1,\dots,u_n) = F(F^{-1}_1(u_1),\dots,F^{-1}_n(u_n))$$

Differentiating Formula \ref{eq:sklar}, a multivariate probability distribution can be constructed by modeling the dependence structure and its one-dimensional marginals separately:

\begin{equation}
\label{eq:joint_f}
f(x_1,\dots,x_n) = c(F_1(x_1),\dots,F_n(x_n)) \cdot f_1(x_1) \cdot \dots \cdot f_n(x_n)
\end{equation}

Where $c$ is the joint copula density function corresponding to the copula function $C$.

 Equation \ref{eq:joint_f} shows that the joint density function $f$ can be expressed as the product of its marginal densities and a dependence structure encoded by $c(F_1(x_1),\dots,F_n(x_n))$.

\vspace{3mm}

In general, a non-Gaussian (for example, Archimedean) multivariate copula function can typically be described using 1, 2 or 3 parameters, which is not flexible enough to describe the multiple kinds of dependences between the marginals in a multivariate random vector. In \cite{bedford1} it was shown that $c(F_1(x_1),\dots,F_n(x_n))$ can be split into a special product (see Equation \ref{joe}), whose elements are pair-copulas and conditional pair-copula p.d.f.'s . This formula was assigned to a specific graph structure, made up of a sequence of trees.

A vine copula, with a general vine structure, is called \textbf{R-vine} (or \textbf{regular vine}), as opposed to a D-vine with trees being chains, or C-vines with trees being stars. 

Researchers working in this field observed that in trees on higher levels, pair copulas gradually get closer to the conditional independence copula. Based on this observation, the important concept of truncated vine copulas was introduced by \cite{brechmann2013risk}: 

A \textbf{truncated vine at level $k$} is a vine copula where all trees at a higher level than $k$ have conditional independent pair-copulas assigned.

\begin{figure}[h]
	\centering
	%\vspace{-1cm}
	\includegraphics[width=0.6\textwidth]{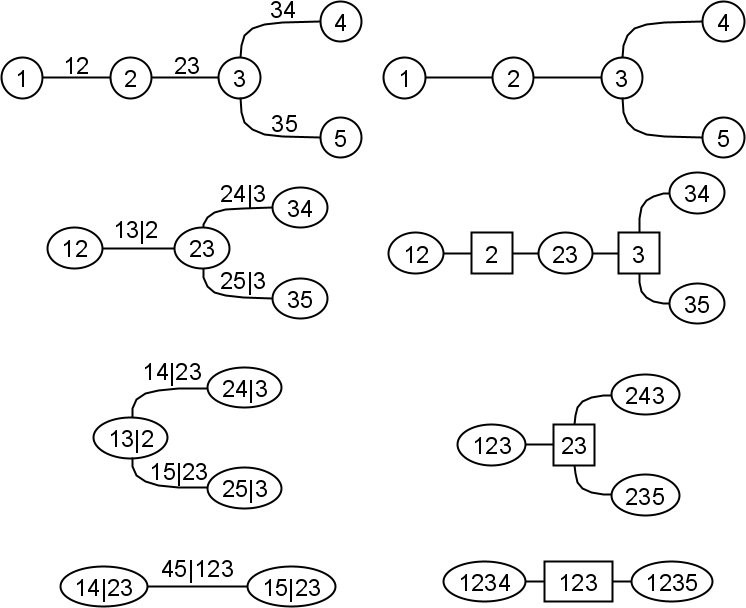}
	\caption{Vine representations: classical (left), cherry (right)}
	\label{cherry_to_vine}
\end{figure}

\vspace{3mm}
Now we recall two graph representations and a matrix representation. The relations between the earlier representations and the newer ones are described in detail in \cite{pfeifer2024vine}. It is important to highlight here that the cherry tree-based representations of \cite{kovacs_szantai} and the corresponding matrix encoding \cite{pfeifer2024vine} are essential to the new approach introduced in the current paper. The original, and most popular representation was introduced in \cite{bedford2}. For a quick look on the relations between them, we give the following short illustrative example to the reader, see Figure \ref{cherry_to_vine}.

\vspace{2mm}

\textbf{Vine representations}

A popular formula of the joint probability density function (\ref{eq:joint_f}) introduced by \cite{bedford2} can be given by

\begin{equation}
\label{joint_f2}
f(\vb{x}) = \prod_{j=1}^n \underbrace{f_j(x_j)}_{\begin{matrix}[0.5]\text{\scriptsize one-dimensional}\\ \text{\scriptsize marginals}\end{matrix}} \prod_{i=1}^{n-1} \prod_{(e_1,e_2|S) \in E_i} \underbrace{c_{e_1,e_2|S}}_{\begin{matrix}[0.5]\text{\scriptsize conditional pair-copulas}\\ \text{\scriptsize in the vine structure}\end{matrix}}\underbrace{(F_{e_1|S}(x_{e_1}|\vb{x}_S),F_{e_2|S}(x_{e_2}|\vb{x}_S))}_{\begin{matrix}[0.5]\text{\scriptsize conditional}\\ \text{\scriptsize c.d.f.'s}\end{matrix}}
\end{equation}

Where $F_{e|S}(x_e|\vb{x}_S)$ is a conditional c.d.f. This is usually not easy to approximate directly from data, however \cite{joe1996families} showed that the conditional c.d.f.'s inside the $c$ function can be calculated the following way:

\begin{equation}
\label{joe}
F_{e|S}(x_e|\vb{x}_S) = \frac{\partial C_{e|S \setminus \{j\}}(F_{e|S \setminus \{j\}}(x_e|\vb{x}_{S \setminus \{j\}}),F_{j|S \setminus \{j\}}(x_{j}|\vb{x}_{S \setminus \{j\}}))}{\partial F_{j|S \setminus \{j\}}(x_{j}|\vb{x}_{S \setminus \{j\}})},
\end{equation}

where $j$ is an element of $S$.

This formula helpful because $F_{e|T \setminus \{j\}}$ and $F_{j|T \setminus \{j\}}$  have already appeared in the previous tree of the vine structure (for certain $j$ values), so the $F$ c.d.f. can be recursively calculated. This was later used by \cite{Dissmann} in the Dißmann algorithm. 

%\begin{definition}
%\label{d_vine}
 %D-vine is a vine structure with the property that in each of the $T_k$ trees, each cluster has the property to be connected to at most $2$ other clusters. 
%\end{definition}

%\begin{wrapfigure}{r}{0.35\textwidth}
%	\centering
 %   \vspace{-0.9cm}
%	\includegraphics[width=0.35\textwidth]{D_vine_example}
%	\caption{Example for a D-vine structure on 5 variables}
%	\label{d_vine}
%\end{wrapfigure}

%After this, a question might arise, namely, what kind of vine structure should we use in Formula (\ref{joint_f2}), which one would be optimal and why?

\vspace{3mm}
%\textbf{Vine defined by a sequence of cherry trees}
% klasszikus, cseresznyefás, (chordal, hivatkozás szinten), previous work, graphical representation reference
% hogyan rendeljük hozzá a kopulákat

Let us recall the vine copula built on a sequence of cherry trees.

According to a paper by \cite{cherry1}, by Definition 4.1, a vine structure is defined the following way using cherry trees. Earlier in Section \ref{section:cherrydefs} we have already seen how a cherry tree corresponds to a cluster tree. Based on this, the vine structure was introduced as a sequence of cherry trees, in \cite{kovacs_szantai}. In this paper, we give this definition in a more accessible manner as follows:

\begin{definition}
\label{cherry_vine_structure}
The cherry-vine graph structure is defined by a sequence of regular cherry trees $T_1, T_2, \ldots, T_{n-1}$ as follows
\begin{itemize}
    \item $T_1$ is a regular tree on $V = \{1, \ldots, n\}$, the set of edges is $E_1 = \{K_i^1 = (l_i, m_i) ,\ i = 1, \ldots, n-1,\ l_i, m_i \in V\}$
    
    \item $T_2$ is the second order cherry junction tree on $V = \{1, \ldots, n\}$, with the set of clusters $\mathcal{K}_2 = \{K_i^2,\ i = 1, \ldots, n-1 \mid K_i^2 = K_i^1\},\ |K_i^1| = 2$
    
    \item $T_r$ is one of the possible $r$ order cherry junction trees on $V = \{1, \ldots, n\}$, with the set of clusters $\mathcal{K}_r = \{K_i^r,\ i = 1, \ldots, n - r + 1\}$, where each $K_i^r,\ |K_i^{r}| = r$, and is obtained from the union of two linked clusters in the $(r-1)$ order cherry junction tree $T_{r-1}$.
\end{itemize}

\end{definition}

Now we will assign the following pair- and conditional pair-copulas to the vine structure obtained this way:

The following copula p.d.f.'s are assigned to each edge of $T_1$:

$$c_{l_i,m_i}(F_{l_i}(x_{l_i}),F_{m_i}(x_{m_i}))$$

For all $T_r$ trees ($r \in \{2,\dots,n-1\}$), for all connected cluster pairs $A$ and $B$ in $T_r$, the pair-copula p.d.f.'s assigned to them is

$$c_{a,b|S}\left(F_{a|S}\left(x_{a}|\boldsymbol{x}_{S}\right),F_{b|S}\left(x_{b}|\boldsymbol{x}_{S}\right)|\boldsymbol{x}_{S}\right)$$

where

$$S = A \cap B, a = A \setminus S, b = B \setminus S$$

Since $A$ and $B$ are connected in $T_r$, their symmetric difference contains two elements, thus the sets $a$ and $b$ contain exactly one element.

According to a paper by \cite{cherry1}, Theorem 4.2, the joint p.d.f. can be expressed the following way:

\begin{equation}\label{eq:cherry_vine_density}
\mkern-18mu f(\boldsymbol{x}) = \left[\prod_{i=1}^n f_i(x_i)\right] \left[\prod_{i=1}^{n-1} c_{K_i^1}\left(F_{l_i}(x_{l_i}),F_{m_i}(x_{m_i})\right)\right] \prod_{\substack{A,B \in \mathcal{K}_r \\ A, B \text{ connected }}} \prod_{\substack{a \in A, b\in B \\ S = A \cap B}} c_{a,b|S}\left(F_{a|S}\left(x_{a}|\boldsymbol{x}_{S}\right),F_{b|S}\left(x_{b}|\boldsymbol{x}_{S}\right)|\boldsymbol{x}_{S}\right)
\end{equation}

where the value of $F_{a|S}$ was defined the following way by \cite{joe1996families}:

$$F_{j|S}(x_j|\boldsymbol{x}_S) = \frac{\partial C_{i,j|S \setminus \{i\}}(u_i,u_j)}{\partial u_i} \Biggr\rvert_{\begin{matrix}[0.5] {\scriptsize u_i = F_{i|S \setminus \{i\}}(x_i|\boldsymbol{x}_{S \setminus \{i\}})} \\ {\scriptsize u_j = F_{j|S \setminus \{j\}}(x_j|\boldsymbol{x}_{S \setminus \{j\}})} \end{matrix}}$$

for all $i \in S$, where $S \subset V$ and $j \in S$.

It follows that when truncating a vine structure on level $t$, the truncated vine copula assigned to it takes the same shape as Formula (\ref{eq:cherry_vine_density}), with conditional independence pair copulas in all trees after the $t$'th tree.

\vspace{3mm}

\textbf{Matrix encoding of a vine structure}

The matrix encoding of a vine structure from a graphical input has a central role in this paper. The initial matrix encoding was introduced by \cite{vinematrix}, to count the number of vine structures, which builds the matrix column-wise. A row-wise encoding as well as relations between the two matrix encodings were introduced in our earlier paper \cite{pfeifer2024vine}. This row-wise encoding is also capable of storing truncated vines directly instead of having to build the full vine, which is the main advantage of this new encoding algorithm. This paper also introduced the concept of "perfect elimination ordering of a vine structure". In the cherry tree representation, all trees of the vine can be transformed into a chordal graph. This ordering is a perfect elimination ordering of all trees of the chordal graph sequence. It was shown by \cite{pfeifer2024vine} that such an ordering exists for any vine structure. The elements in this ordering will show up on the main diagonal of the vine encoding matrix, from the top left to the bottom right corner.

Nápoles' encoding was later used to store vines in a digital environment by \cite{Dissmann}. These matrices are, in all cases, $n \times n$ lower-triangular matrices, and their elements are the numbers $1, \dots, n$ which are the indices of the random variables.

\section{Former results related to cherry tree copulas}

In this section, we recall some important results published earlier, which will be used in the present paper. 
The $t$-order cherry tree copula was introduced in Definition 2.11 of paper \cite{kovacs_szantai} as follows:

\begin{definition}
A cherry tree copula density corresponding to  a cherry tree density (see Definition \ref{def:cherry_tree_density}) is given by the formula:
\begin{equation}\label{cherrytree_cop}
    c^{ch}_{\textbf{X}}(\boldsymbol{u}) = \frac{\prod_{K \in \mathcal{K}} c_{\textbf{X}_K}(\boldsymbol{u}_K)}{\prod_{S \in\mathcal{S}} [c_{\textbf{X}_S}(\boldsymbol{u}_S)]^{\nu_S-1}}
\end{equation}
where $\mathcal{K}$ are the clusters consisting, $\mathcal{S}$ are the separators of the cherry tree, and $\nu_S$ is the number of clusters that contain the separator $S$.
\end{definition}

From the definition of truncated vines follows that after truncation, all the conditional pair copula densities are set to $1$ (they are conditional independence copulas).
 
\vspace{3mm}

A second fundamental result, which can be read in the  \cite{cherry1} and \cite{kovacs_szantai} papers, is the following:

\begin{theorem}
\label{truncated_copula_is_cherry}
A vine-copula truncated at level $t-1$ can be expressed as regular $t$-order cherry tree copula (see Formula \ref{cherrytree_cop}).
\end{theorem}

It is important to note that Theorem \ref{truncated_copula_is_cherry} shows that from an analytical point of view, a truncated vine copula is determined by the first cherry tree after the truncation.
According to this, the approximation given by a truncated copula is actually determined by the post-cut cherry tree copula. Since $c^{ch}_{\boldsymbol{X}}$ is a continuous probability density function, we can apply Formula (\ref{KL_folyt}) to obtain the following theorem:

\begin{theorem}
The KL divergence between a truncated vine copula density determined by Formula \ref{cherrytree_cop} associated to the first tree after truncation, and the real joint copula density is given by:
 \begin{equation}
\label{KL_cherrycop}
\text{KL}(c^{ch}_{\textbf{X}}(\boldsymbol{u}),c_{\textbf{X}}(\boldsymbol{u}))=\sum_{i=1}^n H\left( U_{i}\right) - H\left( 
\textbf{U}\right) - \left(\sum_{K\in \mathcal{K}} I(\textbf{U}_{K}) - \sum_{S\in \mathcal{S}} (\nu_{s}-1) I(\textbf{U}_{S})\right)
\end{equation}

where $\nu_{S}$ the multiplicity of separator $S$, $c_{\mathbf{x}}$ is the joint copula density, $c^{\text{ch}}_{\mathbf{x}}$ is the cherry copula approximation corresponding to its $t$'th tree, $\mathcal{K}$ and $\mathcal{S}$ are the cluster and separator sets of $T_t$, $U_i$ are the uniform marginals and $\mathbf{U}$ is the copula density $c_{\mathbf{x}}$ of $(U_1,...,U_n)$.
\end{theorem}
 
\vspace{2mm}
Let us make the following important remarks:
\begin{itemize}
    \item  $\sum_{i=1}^n H\left( U_{i}\right) =0$, since a uniform distribution's entropy on the unit interval equals 0.
    \item The term $H(\mathbf{U})$ is the copula entropy which equals to $-I(\textbf{X})$ therefore $-H(\mathbf{U})\ge 0$.
\end{itemize}

From Formula (\ref{KL_cherrycop}) it follows that $\text{KL}(c^{ch}_{\textbf{X}}(\boldsymbol{u}),c_{\textbf{X}}(\boldsymbol{u}))$ between the cherry tree copula density which describes the truncated vine and the copula density (or empirical copula data) depends only on the information contents associated to the clusters and separators of the first tree after truncation, by the following expression: 
\begin{equation}
\label{weight of copula dens approx}
    \sum_{K\in \mathcal{K}_{k}} I(\textbf{U}_{K}) - \sum_{S\in S_{k}} (\nu_{s}-1) I(\textbf{U}_{S}).
\end{equation}

In the following for sake of simplicity we use the notation $c_{ch}(u)$ for $c^{ch}_X(u)$ and the abbreviation $c(u)$ instead of $c_x(u)$.

Therefore a very important idea, which is also at the base of our new approach, is that minimising the value of $\text{KL}(c_{{ch}},c)$ at a given level $t$ will give us the best cherry tree copula level $t$. Let us note here, that in order to use this we have to prove that for a cherry tree copula there exists a truncated vine copula whose first tree after truncation is this cherry tree in the. This will be proved in the next section.

\section{Cherry tree copula obtained from a truncated vine}

\subsection{Finding a truncated vine distribution for a given cherry tree distribution}

\label{subsection:truncatedvine}
% Edith + Dani
% saját tétel + példa
% D-vine képlet, claster / separator képlet
% cherry tree distribution = truncated vine (bizonyos esetekben: ha a szeparálók is cherry-t alkotnak)

For the sake of simplicity, in this section, we will denote the $c(F_1(x_1),\dots,F_n(x_n))$ copula density function introduced in Formula (\ref{eq:joint_f}) as $c(1,\dots,n)$.

In order to prove the main theorem of this part we need the following Lemmas:

\begin{lemma}
\label{copula_d_vine}
Any $c$ copula density function on $n$ variables can be expressed as a D-vine. This representation isn't unique, the D-vine formula depends on the permutation of indices $\phi \in \text{Perm}(n)$:

$$c(1,\dots,n) = \prod_{i=1}^{n-1} \prod_{j=1}^{n-i} c(\phi(i),\phi(i+j)|\phi(i+1),\dots,\phi(i+j-1))$$

Where the condition is empty if $i+1 > i+j-1$ or if $j = 1$.
\end{lemma}

% kell: referencia, ahol bevezetik a D-vine-okat

\begin{lemma}
\label{copula_division_d_vine}
The following ratio can be expressed as a product for any $c$ copula density function, for any $\phi \in \text{Perm}(2,\dots,n)$:

$$\frac{c(1,\dots,n)}{c(2,\dots,n)} = \prod_{i=1}^{n-1} c(1,\phi(i)|\phi(1),\dots,\phi(i-1))$$
\end{lemma}

\begin{pf}
We will use Lemma \ref{copula_d_vine} for the nominator and the denominator. Let us take a  permutation from $\text{Perm}(n)$, and let us denote its elements by $1,2,\dots,n$. Then

$$\frac{c(1,\dots,n)}{c(2,\dots,n)} = \frac{\prod_{i=1}^{n-1} \prod_{k=1}^{n-i} c(i,i+k|i+1,\dots,i+k-1)}{\prod_{i=2}^{n-1} \prod_{k=1}^{n-i} c(i,i+k|i+1,\dots,i+k-1)}$$

We have the same formula in the nominator and the denominator except the one in the denominator starts from $i=2$. Every factor that appears in the denominator appears in the nominator as well, so the entire denominator cancels out. Only the $i=1$ case remains, so the first product sign disappears and $i=1$ is substituted.

$$\prod_{k=1}^{n-1} c(1,1+k|1+1,\dots,1+k-1) \stackrel{(*)}{=} \prod_{i=1}^{n-1} c(1,i|1,\dots,i-1)$$

$(*)$ Substituting $i:=k+1$.

Since we have denoted the elements of $\phi$ as $1,\dots,n$, this is the same formula as in the Lemma. So what we have obtained are the product of all factors that contain the index $1$. $\square$
\end{pf}

% Ezt mégse használjuk:

%\begin{lemma}
%\label{num_of_elements_cherry}
%A cherry tree on $n$ variables with $m$ clusters has $n-m+1$ elements in each cluster.
%\end{lemma}

%\begin{pf}

%PAPÍRRA!!

%Let $k :=$ The number of elements in each cluster.
%With $m$ clusters, there are $m-1$ separators. Each separator contains $k-1$ elements.
%We will prove the statement by induction.
%For $m=2$ clusters, there are $n-1 = n-2+1 = n-m+1$ elements in each cluster.
%Assume the Lemma holds for $m$ clusters. Now we will prove the statement for $m+1$ clusters.
%Let us separate one cluster and separator that was newly added to the construction. The cluster contains one new element, and the separator does not. So comparing the $m+1$ cluster case to the $m$ cluster case, we have to have one fewer element in each cluster for there to be $n$ total indices; giving us $(n-m+1)-1= n-(m+1)-1$ elements in each cluster.
%\end{pf}

From the definition of the D-vine and the definition of the truncated vine, it follows that we can define the truncated D-vine distribution:

\begin{definition}
\label{distr_of_truncated_D_vine}
The distribution corresponding to a D-vine on $k$ indices truncated at level $k$ is
$$D_{k}(1,\dots,n) := \prod_{i=1}^{k} \prod_{j=1}^{n-i} c(i,i+j|i+1,\dots,i+j-1) \prod_{i=1}^n f(i)$$
\end{definition}

\begin{rmk}
This is equivalent to using $\phi = (1,\dots,n)$ in Lemma \ref{copula_d_vine}, with $n-1=k$ in the first product.
\end{rmk}

\begin{lemma}\label{two_connections}
Each cluster of a tree in a vine structure is connected to other clusters through at most two distinct separators. (This corresponds to the proximity condition.)
\end{lemma}

\begin{pf}
Reductio ad absurdum: Let us assume that there exists a cluster $(X_1,\dots,X_k)$ in the given tree of the vine structure that connects to $3$ other clusters $(X_2\dots,X_k,X_{k+1}),(X_1,X_3,\dots,X_k,X_{k+2})$ and $(X_1,X_2,X_4,\dots,X_k,X_{k+3})$ through $3$ distinct separators $S_1=(X_2,\dots,X_k),S_2=(X_1,X_3,\dots,X_k)$ and $S_3=(X_1,X_2,X_4,\dots,X_k)$.
However, these new, $k-1$-element clusters do not satisfy the running intersection property in the previous tree. No matter which one is in the middle, there will be one $X_i$ that appears in the left and the right clusters, but not the middle one.
\begin{itemize}
\item When $S_1$ is in the middle, $S_2$ and $S_3$ contain $X_1$, but $S_1$ does not.
\item When $S_2$ is in the middle, $S_1$ and $S_3$ contain $X_2$, but $S_2$ does not.
\item When $S_3$ is in the middle, $S_1$ and $S_2$ contain $X_3$, but $S_3$ does not.
\end{itemize}
$\square$
\end{pf}

\begin{theorem}
\label{truncated_vine_is_cherry}
For all $k$-order regular cherry tree distributions on $n$ variables (where $n>k$) there exists vine on $n$ variables truncated at level $k-1$ such that the joint distribution they describe is identical.
\end{theorem}

\begin{pf}

We will prove the statement by induction for $n$, such that $n>k$.

\underline{$n=k+1$ case:}

Without loss of generality, let us assume that the elements are $\{1,\dots,n\}$. Then their joint p.d.f. can be written up the following way due to the conditional independence $1,k+1|2,\dots,k$ the cherry tree describes:

$$f(1,\dots,k+1) = \frac{f(1,\dots,k)\color{blue}f(2,\dots,k+1)}{\color{blue}f(2,\dots,k)} = (*_1)$$

Using Formula (\ref{eq:cherry_vine_density}) for all $f$'s with the simplified notation introduced at the beginning of this section:

$$(*_1) = \frac{c(1,\dots,k)\color{blue}c(2,\dots,k+1)}{\color{blue}c(2,\dots,k)} \prod_{i=1}^n f(i) = (*_2)$$

Using Lemma \ref{copula_division_d_vine} for $\color{blue}\frac{c(2,\dots,k+1)}{c(2,\dots,k)}$ with $\phi = (2,\dots,k+1)$:

$$(*_2) =  c(1,\dots,k) \color{blue} c(k+1,k) c(k+1,k-1|k) \dots c(k+1,2|3,\dots,k) \color{black} \prod_{i=1}^{k+1} f(i) = (*_3)$$

Using Lemma \ref{copula_d_vine} for $c(1,\dots,k)$ with $\phi = (1,\dots,k)$:

\begin{align*}
&(*_3) = c(1,2) c(2,3) \dots c(k,k+1) \color{blue} c(k,k+1) \color{black} \\
&c(1,3|2) c(2,4|3) \dots c(k-2,k|k-1) \color{blue} c(k-2,k+1|k-1) \color{black} \\
&c(1,4|2,3) c(2,5|2,4) \dots c(k-3,k|k-1,k-2) \color{blue} c(k-3,k+1|k,k-1) \color{black} \\
&\vdots \\
&c(1,k|2,\dots,k-1) \color{blue} c(2,k+1|3,\dots,k) \color{black} \\
&\prod_{i=1}^{k+1} f(i) = (*_4)
\end{align*}

With this we have proved the $n=k+1$ case.

Now we suppose the induction step, which is that the cherry tree density function on $n$ variables $X_1,...,X_n$ can be written as a truncated vine. We will prove that a $k$-width cherry tree copula on $n+1$ variables $X_1,...,X_n,X_{n+1}$ is also written as a truncated vine.
Due to the hypothesis we have
\begin{equation}
\label{joint_f2_cherry}
f(\vb{x}) = \prod_{j=1}^n {f_j(x_j)} \prod_{i=1}^{n-k+1}\prod_{(e_1,e_2|T) \in E_i} {c_{e_1,e_2|T}}{(F_{e_1|T}(x_{e_1}|\vb{x}_T),F_{e_2|T}(x_{e_2}|\vb{x}_T))}
\end{equation}
First, let us see how the new $(n+1)$st variable can be added to a $k$-width regular cherry tree.
\begin{enumerate}
    \item by using one of the existing separators
    \item by using a new separator which is a subset of an existing cluster
\end{enumerate}
The case of an existing separator.
In this case the formula above is multiplied by
$$\frac{\color{blue}f(S,n+1)}{\color{blue}f(S)}f(n+1) = (*_5)$$

1. Let us suppose that the separator $S=\{\phi(1),\dots,\phi(k-1)\}$ $(\phi \in \text{Perm}(n))$ such that $S$ is an existing element of the set of separators.
 
\begin{equation*}(*_5) =\frac{c(\phi(1),\dots,\phi(k-1),n+1)}{c(\phi(1),\dots,\phi(k-1))}f(n+1) \stackrel{\text{Lemma } \ref{copula_division_d_vine}}{=} \prod_{i=1}^{n-1} c(n+1,\psi(i)|\psi(1),\dots,\psi(i-1))f(n+1)\end{equation*}

%% A \psi definícióját jobban leírni

where the permutation $\psi \in \text{Perm}(\phi(1),\dots,\phi(k-1))$ is chosen such that $\forall i \in \{1,\dots,k-2\}$ the separator S in the formula $\{\psi(1),\dots,\psi(i-1)\}$ appears in $T_i$. This can be chosen because these separators form a sequence of inclusive sets due to the definition of the vine structure, i.e. for example

$$\{\psi(2)\}\subseteq\{\psi(2),\psi(7)\} \subseteq \{\psi(2),\psi(4),\psi(7)\} \subseteq\dots \subseteq \{\psi(1),\dots,\psi(i-1)\}$$

This way the regular cherry tree structure will be preserved.

By multiplying the (conditional) pair copula formula with this new factor, we get the formula for the $n+1$ case:

$$\frac{\prod_{C \in \mathcal{K}_n} c(C)}{\prod_{S \in \mathcal{S}_n} c(S)} \prod_{i=1}^n f(i) \cdot {\color{blue} \prod_{i=1}^{n-1} c(n+1,\psi(i)|\psi(1),\dots,\psi(i-1))} f(n+1) = \frac{\prod_{C \in \mathcal{K}_{n+1}} c(C)}{\prod_{S \in \mathcal{S}_{n+1}} c(S)_n} \prod_{i=1}^{n+1} f(i) $$

Where $C_n$ and $S_n$ are the set of clusters and set of separators of all trees on $n$ variables, and $C_{n+1}$ and $S_{n+1}$ are the set of clusters and set of separators of all trees on $n+1$ variables.

This is because the products of (conditional) pair copulas will be extended by one factor on each level, like in the $n=k+1$ case (where a level is defined by the number of variables in the conditioning set, which is a row of conditional pair copula functions in the $n=k+1$ case).

This completes the proof for this case.

\vspace{3mm}

2. Let us suppose that the separator $S=\{\phi(1),\dots,\phi(k-1)\}$ is not an existing element of the set of separators, but it is a potential separator. In order to preserve the vine structure, all of whose trees are regular cherry trees (as seen in Lemma \ref{two_connections}), we also need to make sure that the new cluster we are adding preserves this regularity condition on each level.

In this case we suppose that the new separator does not belong to the existing separators, but belongs to the potential separators. This implies that the new separator is a subset of a cluster that contains a simplicial vertex (a vertex that appears in only one cluster). Otherwise the new cluster would be connected to an existing separator.
Let us denote the cluster containing the simplicial vertex $\phi (k)$ by $%
\left\{ \phi (1)\ldots \phi (k-1)\phi (k)\right\} $. It is clear
that this is a $k$-element cluster in the $k$ order cherry tree. Then we add the ($%
n+1$)st vertex. The cluster $%
\left\{ \phi (1)\ldots \phi (k-1)\phi (k)\right\} $ is connected to
the rest of the existing tree with the separator $\left\{ \phi (1)\ldots
\phi (k-1)\right\} $. The new vertex $n+1$ is connected to the tree with a
new separator which contains $\phi (k)$. Without loss of generality we denote this separator by $%
\left\{ \phi (2)\ldots \phi (k)\right\} $.

We make a step backward (to a lower level tree) in the following way. By Definition \ref{cherry_vine_structure} the two separators will be two clusters in the $k-1$ order regular cherry tree. The new vertex $n+1$ will be connected to the rest of the tree by the intersection between $\left\{\phi(1),\ldots,\phi(k-1)\right\}$ and  $\left\{\phi(2),\ldots,\phi(k)\right\}$, that
is $\left\{\phi(2),\ldots,\phi(k-1)\right\} $. This way the proof is reduced to the former case. $\square$
\end{pf}

\begin{figure}[h]
    \centering\includegraphics[width=0.8\textwidth]{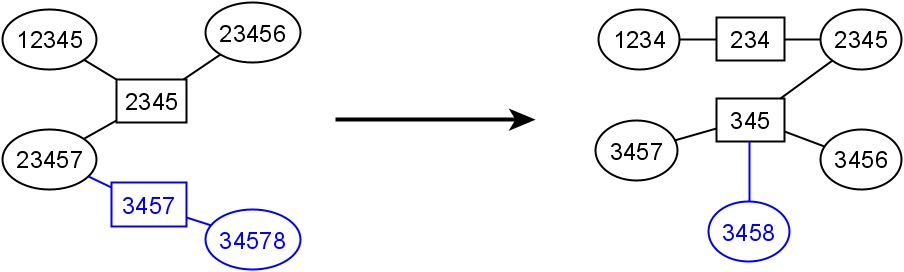}
    \caption{Example for the last step of the proof: Making a step backward in a cherry tree to find that the newly added cluster $34578$ that connects with a new separator $3457$ can be connected with an existing separator in a lower level tree.}
    \label{fig:cherry_to_lower_level}
\end{figure}

\newpage

\begin{ex}\label{trunc_vine_to_cherry_ex}
Writing up a truncated vine corresponding to a given cherry tree.
\end{ex}

Let us start from the cherry tree depicted on the following figure. Its corresponding probability distribution is:

\begin{minipage}{0.5\textwidth}
\begin{align*}
&f(12345678) = \frac{f(12345)f(23456)f(23457)f(34578)}{f(2345)^2 f(3457)} = \\ = \ &\frac{{\color{ForestGreen} c(12345)}{\color{RawSienna} c(23456)}{\color{RoyalBlue} c(23457)}{\color{RedOrange} c(34578)}}{{\color{RawSienna} c(2345)}{\color{RoyalBlue} c(2345)}{\color{RedOrange} c(3457)}} \prod_{i=1}^8 f(i) \stackrel{(*)}{=} \\ = \ &{\color{ForestGreen} c(12)c(23)c(34)c(45)} \\ &{\color{ForestGreen} c(13|2)c(24|3)c(35|4)} \\ &{\color{ForestGreen} c(14|23)c(25|34)} \\ &{\color{ForestGreen} c(15|234)} \\ 
&{\color{RawSienna} c(26|345)c(36|45)c(46|5)c(56)} \\
&{\color{RoyalBlue} c(27|345)c(37|45)c(47|5)c(57)} \\
&{\color{RedOrange} c(38|457)c(48|57)c(58|7)c(78)} \prod_{i=1}^8 f(i)
\end{align*}
\end{minipage}%
\begin{minipage}{0.5\textwidth}

\centering
\includegraphics[width=0.5\textwidth]{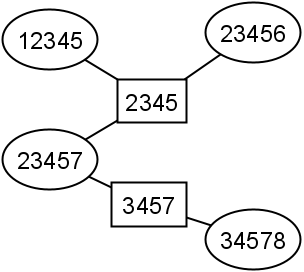}
\\ A cherry tree to convert to a truncated vine
\label{fig:cherry_to_dvine}

\end{minipage}

Where in $(*)$, we have used Lemma \ref{copula_d_vine} for $\color{ForestGreen} c(12345)$, and Lemma \ref{copula_division_d_vine} for $\color{RawSienna} \frac{c(23456)}{c(2345)}$, $\color{RoyalBlue} \frac{c(23457)}{c(2345)}$ and $\color{RedOrange} \frac{c(34578)}{c(3457)}$, respectively.

\begin{figure}[h]
    \centering\includegraphics[width=0.9\textwidth]{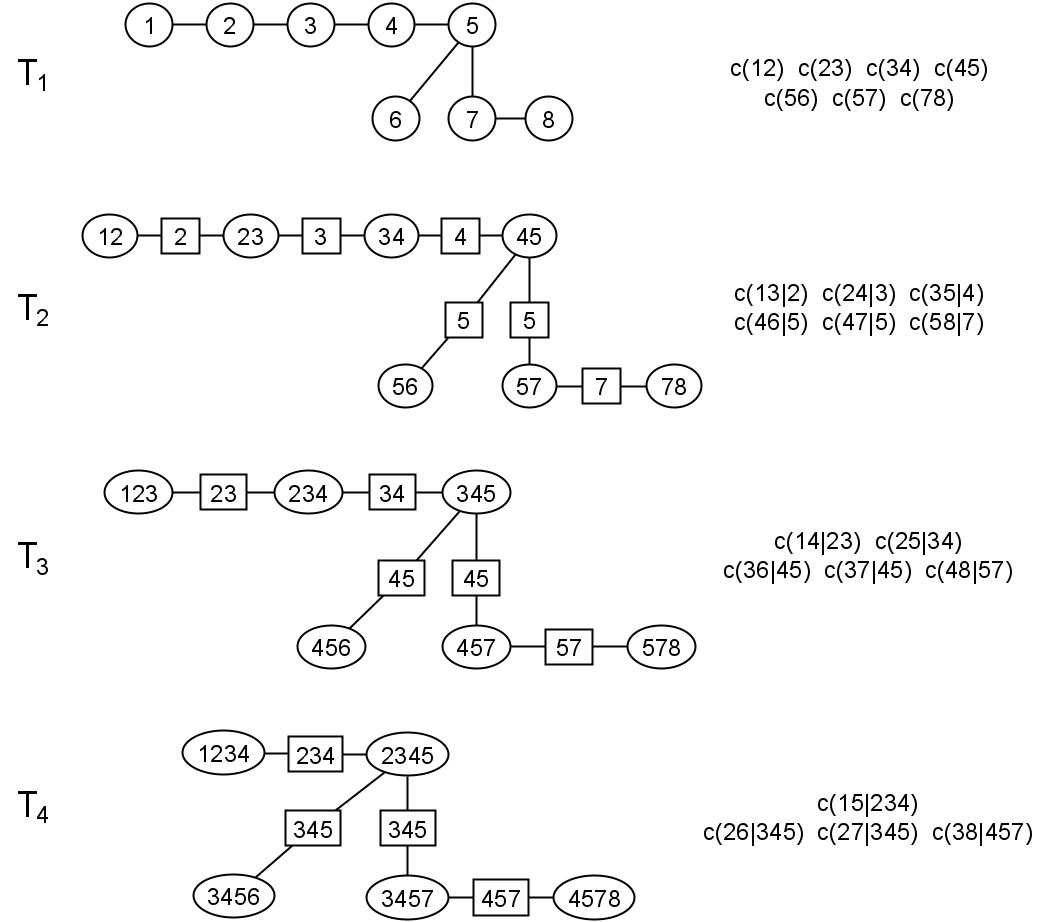}
    \caption{The cherry tree in Example \ref{trunc_vine_to_cherry_ex} can be transformed into this truncated vine, such that the joint copula density they define are the same.}
    \label{fig:cherry_to_trunc_vine_example_constr}
\end{figure}

What we have obtained are exactly the (conditional) pair-copula functions defined by the following truncated vine depicted on Figure \ref{fig:cherry_to_trunc_vine_example_constr}.

\subsection{Weight of a truncated vine}

In the previous Section \ref{subsection:truncatedvine} we have proved that for each regular cherry tree copula, there exists a truncated vine whose first tree after truncation is the given regular cherry.
Formula (\ref{KL_cherrycop}) shows that the KL divergence of a cherry copula tree approximation only depends on the Information contents of the clusters and separators. From these two assertions it follows straightforwardly that the KL divergence of an approximation of a given truncated vine can also be expressed by the same Formula (\ref{KL_cherrycop}).

\begin{definition}
The weight of a truncated vine at level $t-1$, which is defined by the first cherry tree after truncation $(V,\mathcal{K}, \mathcal{S}, \mathcal{M})$ is given by:
\begin{equation}
\label{weight trunc vine}
w(c^{ch}_t) := \sum_{K\in \mathcal{K}_{t}} I(\textbf{U}_{K}) - \sum_{S\in S_{t}} (\nu_{s}-1) I(\textbf{U}_{S}), 
\end{equation}
where $c^{ch}_t$ denotes the truncated vine copula density at level $t-1$, and $\mathcal{K}_t$ denotes the set of clusters, $\mathcal{S}_t$ denotes the set of separators at level $t$.
\end{definition}

Now, after we have proven that from a given regular cherry tree we can theoretically find a truncated vine structure that describes the same copula density, we will also practically find this structure with an algorithm in the next section. This algorithms starts from a given dataset, and it aims to find an optimal truncated vine in a greedy manner, which maximizes the weight defined in Formula (\ref{weight trunc vine}).

%\newpage
\section{Trunc-Opt method for constructing truncated vines}\label{section:greedyalg}
In this section, we aim to find the best-fitting truncated vine at level $t$ for a given dataset. Unlike the traditional methods like  \cite{Dissmann2013selecting} and \cite{brechmann2013risk} algorithms, which find the structure in the lower trees such that they capture the largest dependences, our algorithm's idea is to achieve the best fitting cherry tree at level $t$, by exploiting conditional independences. 

The TruncOpt approach will consist of Algorithm \ref{alg:osnm}, \ref{alg:knc}, and \ref{alg:vm3}.

We are given a dataset of $m$ rows (samples) and $n$ columns (variables), which we will denote by $\mathbf{D}\in\mathbb{R}^{m\times n}$. Let us denote the columns (variables) of this dataset with $\mathbf{X}=(X_1,\dots,X_n)$.

In the following sections, we will work with a Pseudo-Observed version of this dataset, which can be calculated with the R pobs function (see \cite{pobs}). This creates pseudo-observations -- each unique value will be represented by a unique integer, then the whole dataset is divided by the number of rows $m$, which transforms each margin into a uniform distribution. Let us denote the variables obtained after the pobs function by $\mathbf{U}=(U_1,\dots,U_n)$.

Furthermore, we will frequently take a $d$-dimensional subset of the $U_1,\dots,U_n$ variables (where $d \in \{1,\dots,n\}$), which we will simply denote as $\mathbf{U}_D=(U_{i_1},\dots,U_{i_d})$, and we will denote the $j$'th row of the dataset consisting of the columns of $\mathbf{U}_D$ by $U_D^j$. If $D$ is not specified, then $U^j$ denotes the $j$'th row of dataset $\boldsymbol{U}$.

\subsection{Estimating Information Content}

In this section, we will show a methodology for estimating the $d$-dimensional Information Content. For simplicity's sake, we will use the following notation:

$$I(U)=I(U_1,\dots,U_d)$$

\subsubsection{Information content estimation}

We are using the fact that information content of $U=(U_1,\dots,U_d)$ can be expressed as the Kullback-Leibler divergence between a probability distribution $X$ and an independence distribution with the same margins (see Section \ref{subsection:info}).

We have chosen the distribution representing independence to be the $d$-dimensional uniform distribution $Z\sim \text{UNIF}([0,1]^d)$.

% BHOSVYZ betűk szabadok
% E nem jó, mert az az Edge set

A paper by \cite{wang2009divergence} suggests using a k-NN estimation for $\text{KL}(U,Z)$. Namely, let the distance of $U^i$ to its $k$'th nearest uniform neighbor $Z^j$ be denoted by $\nu_k(i)$, and the distance to its $k$'th nearest neighbor in $\{U^j\}_{j\ne i}$ be $\rho_k(i)$. Then the proposed estimator for $\text{KL}(U,Z)$ (with our notations) is

\begin{equation}\label{eq:infKL}
I(U) \approx I_m(U) := \widehat{\text{KL}}_m(U,Z) = \frac{d}{m}\sum_{i=1}^m \log_2\left(\frac{\nu_k(i)}{\rho_k(i)}\right) + \log_2\left(\frac{m}{m-1}\right)  
\end{equation}

where we will be using the choice of $k=5$ neighbors.

An important result adapted from  Theorem 1 in \cite{wang2009divergence}, is the following. If the probability density functions are 1-regular (see Definition 1 \cite{wang2009divergence}), then the divergence estimator, which in our case is the information content estimator, is asymptotically unbiased
\[
\underset{m\rightarrow \infty }{\lim }E(I_{m}\left( U\right)
)=I(U)
\]

\subsubsection{Uniform grid generation}

The above formula will be used several times by the upcoming algorithm. This formula highly depends on the input uniform independent dataset $Z$. To avoid inconsistency, we have decided to not regenerate this dataset $Z$ every time we are calculating a new Information Content. Instead, use the same $d$-dimensional uniform independent dataset, and we have decided to use a grid for that. The coordinates of the points inside a $d$-dimensional grid over $[0,1]^d$ with approximately $m$ points are all $d$-element permutations of

$$\frac{k}{\lceil \sqrt[d]{m}\ \rceil-1 }, \quad k \in \left\{0,\dots,\lceil \sqrt[d]{m}\ \rceil-1\right\}$$

The exact number of points inside this grid will be $\lceil \sqrt[d]{m}\ \rceil^d$, which is comparable to $m$, so $U$ and $Z$ will have a similar number of samples.

Note that ceiling is used instead of floor, since if $\sqrt[d]{m} < 2$, then in the case of floor, we would get a $0$ denominator.

\subsubsection{K-D Tree distance calculation}

% https://dl.acm.org/doi/10.1145/361002.361007

To make the distance calculations faster, we have used a K-D Tree (see \cite{bentley1975multidimensional}) for both the input data $U$ and the grid $Z$. The K-D Tree of $U$ is recalculated every time, because it will always contain a different subset of columns from the total dataset $\mathbf{U} = (U_1,\dots,U_n)$. However we are able to precalculate the K-D Trees of all grids from a $2$-dimensional all the way to a $t$-dimensional grid, where $t$ will be the truncation level of the output vine. This methodology produced a large speed-up, as setting up a K-D Tree is faster than calculating the distances between all pairs $U^i$, $U^j$, after which calculating the distances to the $5$'th nearest neighbors occurred almost instantly, in $O(\log(m))$ time.

\subsubsection{$k$-NN bias reduction}

Finally, in the upcoming algorithm, we will need to compare information contents with a different number of arguments, for example, $I(U_1,U_2)$ to $I(U_1,U_2,U_3)$, where $(U_1,U_2)$ and $(U_1,U_2,U_3)$ are $2$ and $3$ dimensional marginals of $U$. $\widehat{KL}(U,Z)$ introduces a bias based on the dimensionality of $U$ and $Z$. Namely, the larger the dimension, the farther away two neighboring points will be. In expectation, this bias will be the following: If on average, the coordinates of two neighboring points are distance $s$ apart, then in $d$ dimension we can expect for them to be distance $s\sqrt{d}$ apart (due to Pythagoas). Therefore when comparing information contents between dimensions $d$ and $d+1$, we can expect the average distance between two neighboring points to have increased by a factor of $\frac{s\sqrt{d+1}}{s\sqrt{d}}=\frac{\sqrt{d+1}}{\sqrt{d}}$.

To counteract this, we will reduce the density of the smaller dataset by a factor of $\frac{\sqrt{d}}{\sqrt{d+1}}$, by randomly omitting $1-\frac{\sqrt{d}}{\sqrt{d+1}}$ proportion of the data. The upcoming algorithm will compare adjacent information contents, so for the highest required dimension $t$ (which is the truncation level of the output vine), we will not make the data any sparser, then we will omit $1-\frac{\sqrt{t-1}}{\sqrt{t}}$ proportion of it, then $1-\frac{\sqrt{t-1}}{\sqrt{t}}\frac{\sqrt{t-2}}{\sqrt{t-1}} = 1-\frac{\sqrt{t-2}}{\sqrt{t}}$ proportion of it, and so on. So if the highest dimension is $t$, then calculating $I(U_1,\dots,U_d)$ will use a dataset such that $1-\sqrt{\frac{d}{t}}$ elements are randomly discarded.

\subsubsection{Overall steps of Information Content estimation}

\begin{itemize}
\item We are first using the R pobs function on the input columns $(X_1,\dots,X_n)$ to convert it into $[0,1]^n$-valued dataset $(U_1,\dots,U_n)$.
\item We are estimating $I(U) = I(U_1,\dots,U_d)$ with the $k$-NN estimator of $\widehat{KL}(U,Z)$ (with $k=5$), where $Z$ is a grid on $[0,1]^d$ of $\lceil \sqrt[d] {m}\ \rceil^d$ elements.
\item We are using two K-D Trees to calculate $\nu_k(i)$ and $\rho_k(i)$ in the formula of $\widehat{KL}(U,Z)$ significantly faster.
\item The K-D Trees of all grids from dimensions $2$ to $t$ (truncation level) are precalculated, while the K-D Trees of the $U$ subset of columns of $\mathbf{U}=(U_1,\dots,U_n)$ are calculated when necessary.
\item For each level $d\in\{2,\dots,t\}$, we are randomly discarding $1-\sqrt{\frac{d}{t}}$ proportion of the rows of the input dataset $U$ before the calculation to counteract the $k$-NN bias.
\end{itemize}

\subsection{The proposed truncated vine building algorithm}

In this section, we will present our new cherry tree-based truncated vine building methodology. However, first we need a small helper function (Algorithm \ref{alg:osnm}) that will be called by the main method, Algorithm \ref{alg:knc}.

This helper function calculates the one-step neighborhood matrix $A$ of a given input graph $G=(V,E)$. The elements of this matrix are all lists of size $L-2$ ($L$ will be a level defined later in Algorithm \ref{alg:knc}). Element $(i,j)$ of the output matrix lists all the vertices $v\in V$ such that there is a path $(i,v,j)$ in $G$. If there are exactly $L-2$ of such vertices $v$, then in Algorithm \ref{alg:knc} we will be able to find a clique of size $L$ in the specific cherry tree structure it is working with.

\vspace{3mm}
\begin{breakablealgorithm}
\caption{One-step neighborhood matrix}\label{alg:osnm}
\hspace{-6mm}\textbf{Input:} Graph $G=(V,E)$ on $|V|=n$ vertices, current level $L \ge 3$.\begin{algorithmic}[1]
\State Let $A$ be an empty $n \times n$ matrix.
\For{$s,t \in V$}
    \State $W = \text{Ne}(s) \cap \text{Ne}(t) \setminus  \{s\} \setminus \{t\}$
    \If{$|W| = L-2$}
        \State $A_{st} = W$
    \EndIf
\EndFor
\end{algorithmic}
\textbf{Output:} $A$, the one-step neighborhood matrix.
\vspace{1mm}
\end{breakablealgorithm}

\vspace{3mm}

Now we will present the main algorithm of the paper, a new truncated vine building methodology, based on cherry trees. We will use the notations introduced in Section \ref{section:greedyalg}.

\vspace{3mm}

\begin{breakablealgorithm}
\caption{$t$-neighborhood cherry}\label{alg:knc}
\hspace{-6mm}\textbf{Input:} The $\boldsymbol{D} \in \mathbb{R}^{m \times n}$ dataset, $t \in \{2,\dots,n\}$ truncation level, and $I: \mathbb{R}^{m \times d} \to \mathbb{R}$ information content functions (for any $d \in \{2,\dots,n\}$) estimated from the $\boldsymbol{D}$ dataset.
\begin{algorithmic}[1]
\State Let $M$ be the matrix of Mutual Informations: $M_{ij} := I(U_i,U_j)$, $i,j \in \{1,\dots,n\}$.
\State Let $\Gamma$ be a complete graph on $n$ vertices, with edge weights $M_{ij}$, $i,j \in \{1,\dots,n\}$.
\State Let $G_2$ be a maximum weight spanning tree of $\Gamma$.
\State $w_2 := \sum_{(i,j)\in E(G_2)} I(U_i,U_j)$.
\State $\mathcal{K}_2 := E(G_2)$

\For{$L=3$ to $t$}
	\State Let $F = $ [ ] be an empty list of the found indices.
    \State Let $\mathcal{K}_L = $ [ ] be an empty list of the found clusters.
	\State Let $E = $ [ ] be an empty list of the found edges.
    \State $G_L := G_{L-1}$
    \State $w_L := 0$
    \State Calculate $A$, the one-step neighborhood matrix of $G_L$ on level $L$.
    \For{$r=1$ to $n-t+1$}
        \State Let $P = $ [ ] be an empty list containing all potential clusters.
        \If{$F$ is an empty list}
            \State $V := $ [$1,\dots,n$]
        \Else
            \State $V := F$
        \EndIf
        \For{$i \in V$, $j=1$ to $n$}
            \If{$A_{ij}$ exists}
                \State Add the list [$i$, all elements of $A_{ij}$, $j$] to $P$.
            \EndIf
        \EndFor
        \If{$r=1$}
            \State $w_L \pluseq \max\left\{I(\boldsymbol{U}_K) | K\in P \setminus \mathcal{K}_L\right\}$
            \State $Q := \arg\max\left\{I(\boldsymbol{U}_K) | K\in P \setminus \mathcal{K}_L\right\}$
        \Else
            \State $w_L \pluseq \max\left\{I(\boldsymbol{U}_{(i,A_{ij},j)})-I(\boldsymbol{U}_{(i,A_{ij})}) | (i,A_{ij},j)\in P \setminus \mathcal{K}_L\right\}$
            \State $Q := \arg\max\left\{I(\boldsymbol{U}_{(i,A_{ij},j)})-I(\boldsymbol{U}_{(i,A_{ij})} | (i,A_{ij},j)\in P \setminus \mathcal{K}_L\right\}$
        \EndIf
        \State Add $Q$ to $\mathcal{K}_L$.
        \State Add $i,j,$ and any element of $A_{ij}$ to $F$ that $F$ does not contain yet.
        \State Using $Q = (i,A_{ij},j)$, add the $(i,j)$ edge to $E$.
    \EndFor
    \State Add all edges in $E$ to $G_L$.
\EndFor
\end{algorithmic}
\textbf{Output:} $G_2,G_3,\dots,G_t$ cherry tree graphs, $\mathcal{K}_2,\mathcal{K}_3,\dots,\mathcal{K}_t$, their maximum clique clusters, and $w_2,w_3,\dots,w_t$, the weight of each cherry tree.
\vspace{1mm}
\end{breakablealgorithm}

\vspace{3mm}

In order for the above method to be usable by modern software such as the R VineCopula package of \cite{rvine}. To be comparable to other similar methods, we will need to encode the output cherry tree sequence into a vine matrix introduced by \cite{vinematrix}. For doing this we have to prove that the output of the above procedure is a regular cherry tree. We will prove this in Section \ref{section:regularity_proof}.

The following algorithm is capable of the vine matrix encoding, which now algorithmically shows that regular cherry tree sequences are indeed truncated vines (see Section \ref{section:regularity_proof} for a mathematical proof).

\vspace{3mm}

\begin{breakablealgorithm}
\caption{Regular cherry to truncated vine matrix encoding}\label{alg:vm3}
\hspace{-6mm}\textbf{Input:} $\mathcal{K}_2,\mathcal{K}_3,\dots,\mathcal{K}_t$ the maximum clique clusters of a sequence of cherry trees of order $2,3,\dots,t$, $n$: the number of vertices in each cherry tree, trunc: If True, return a truncated vine (truncated to level $t$), otherwise, return a full vine.
\begin{algorithmic}[1]
    \State Let $M$ be an empty $n \times n$ matrix.
    \State Find a vertex in the clusters of $\mathcal{K}_t$ that only appears once in $\mathcal{K}_t$ (a leaf node), and call it $v$.
    \State Let us find and store the reversed perfect elimination ordering of the input cherry trees in $P = $ [$v$]. This will be a perfect elimination ordering of all cherry trees in the input sequence. To see why this is, please refer to \cite{pfeifer2024vine}.
    \For{$i=2$ to $t$}
        \State Find a cluster $K \in \mathcal{K}_i$ that contains all elements of $P$.
        \State Add the single element of $K \setminus P$ to $P$.\Comment{Remark \ref{rmk:k_p}}
    \EndFor
    \For{$i=1$ to $t$}
        \State $M_{n-i+1,n-i+1} := P$[$i$]
    \EndFor
    \State $M_{n,n-1} := v$
    \State $S :=$ [$P$[$1$],$P$[$2$]]
    \For{$i=1$ to $t-2$}
        \State $R :=$ [ ], the clusters which will be represented in the current column (one of the last $t$ columns)
        \For{$j=i+2$ to $2$}
            \For{$K \in \mathcal{K}_j$}
                \If{$M_{n-i-1,n-i-1} \in K$ and if all elements of $K \setminus \{M_{n-i-1,n-i-1}\}$ are in $S$}
                    \State Add $K$ to $R$.
                    \State Break out of the innermost for loop.
                \EndIf
            \EndFor
            \State Add $M_{n-i-1,n-i-1}$ to $S$.
            \State Add [$M_{n-i-1,n-i-1}$] to $R$.
            \For{$r=1$ to $i+1$}
                \State $M_{n-i-1+r, n-i-1} :=$ The single element of $R$[$r$] $\setminus$ $R$[$r+1$]\Comment{Remark \ref{rmk:r_r1}}
            \EndFor
        \EndFor
    \EndFor
    \For{$i=n-t$ to $1$}
        \For{$K \in \mathcal{K}_t$}
            \If{$|K \cap S| = t-1$}
                \State $M_{i,i} :=$ The single element of $K \setminus S$
                \State $R := $ [$K$], the clusters represented in the current column (one of the first $n-t$ columns)
                \State Break out of the innermost for loop.
            \EndIf
        \EndFor
        \For{$j=t$ to $2$}
            \For{$K \in \mathcal{K}_j$}
                \If{$M_{i,i} \in K$ and all elements of $K \setminus \{M_{i,i}\}$ are in $R$[$-1$] (the last cluster added to $R$)}
                    \State Add $K$ to $R$ if $R$ is not in $K$ already.
                    \State Break out of the innermost for loop.
                \EndIf
            \EndFor
        \EndFor
        \State Add $M_{i,i}$ to $S$.
        \State Add [$M_{i,i}$] to $R$.
        \For{$r=1$ to $k-1$}
            \State $M_{n-k+1+r, i} :=$ The single element of $R$[$r$] $\setminus$ $R$[$r+1$]\Comment{Remark \ref{rmk:r_r1}}
        \EndFor
    \EndFor
    
    \If{trunc = False}
        \State Fill out the rest of the matrix (columns $i=n-t$ to $1$, rows $j=i+1$ to $n-t+1$) using Algorithm 2 of \cite{pfeifer2024vine} (Page 14, Example on Page 16-17).
    \EndIf
\end{algorithmic}

\textbf{Output:} The $M$ vine matrix that encodes the input cherry tree.
\end{breakablealgorithm}

\begin{rmk}\label{rmk:k_p}
At this point, $|K\setminus P| = 1$. $P$ starts off with $1$ element, $v$, and $K_2$ has $2$ elements. Generally, $K_i$ is a cluster of size $i$, so it will have $i$ elements.
In Row $6$, the size of $P$ increases by $1$, so in every iteration $i$, $K_i$ will have $1$ more element than $P$, so if $P \subset K$, then $|K\setminus P| = 1$.
\end{rmk}

\begin{rmk}\label{rmk:r_r1}
Generally in vines and cherry tree sequences, the clusters of level $r$ will be the separators of level $r+1$. $R$ contains a chain of such clusters, where cluster $r$ embedded into cluster $r+1$, from $r=i+2, i+1, \dots, 2$. Therefore, $R[r] \setminus R[r+1]$ is a subtraction between sets whose size only differs by $1$, where $R[r+1] \subset R[r]$. Therefore, $|R[r] \setminus R[r+1]| = 1$.
\end{rmk}

\begin{figure}[h!]
    \centering
    \includegraphics[width=0.45\linewidth]{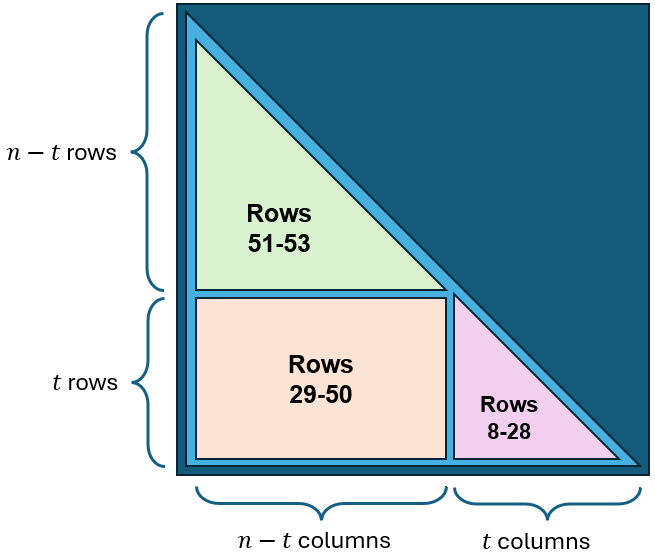}
    \caption{The matrix filling methodology of Algorithm \ref{alg:vm3}}
    \label{fig:placeholder}
\end{figure}

\subsection{The $t$-neighborhood cherry algorithm builds a regular cherry}\label{section:regularity_proof}

\begin{definition}
The one-step neighborhood matrix of an $L$-order regular cherry tree is a matrix whose $(i,j)$ element is the separator-set $S$ between two connected clusters $(i,S)$ and $(S,j)$ in the $L$-order regular cherry tree, containing $L-1$ elements. We say that the vertices $i$ and $j$ are one-neighbor apart from each other, and the clusters $(i,S)$ and $(S,j)$ are called neighbor clusters in the $L$-order regular cherry tree. 
\end{definition}

\begin{theorem}
Algorithm \ref{alg:knc}, which uses a one-step neighborhood matrix, outputs a regular cherry tree in every iteration, i.e. constructing an $L+1$-order regular cherry tree from an $L$-order regular cherry tree.
\end{theorem}

\begin{pf}
Constructing an $L+1$-order cherry tree from an $L$-order cherry tree, which is given as a one-step neighborhood matrix, is done step by step.

The proof uses induction.

\textbf{Verification step:} The input is a tree encoded in a one-step neighborhood matrix and the output is a cherry tree of order $3.$ Note that the one-step neighborhood matrix contains only one element sets in this case and defines the neighbor clusters of a $2$-order cherry tree.

We want to prove that each cluster of the output $3$-order cherry tree is connected to other clusters through at most two separators (this is the definition of regular cherry tree, see in Section \ref{section:cherrydefs}).

The first cluster is formed by finding two vertices which are $1$-neighbors in the one-step neighborhood matrix. This will define two $2$-element neighboring clusters (in the 2-order cherry tree), $(v_1,w)$ and $(w,v_2)$. We join these two clusters and obtain a $3$-element cluster $C = (v_1,w,v_2)$. Let us call the edges of $G_2$ prior subsets. The cluster $C$ obtained in this way will contain two prior subsets which are the joint clusters of 2-order cherry tree.

By this procedure, each new $3$-order cluster and the already connected $3$-order cluster have a prior $2$-element subset in common. But because each cluster has no more than two prior subsets, each cluster is separated from its neighboring clusters by at most two separators, the prior subsets.

\textbf{Induction step:} The input is an $L$-order cherry tree satisfying the regularity condition: each of its clusters is connected to other clusters by at most two separators. The input in this case is a one-step neighborhood matrix whose elements are the separator sets of the $L$-order cherry
tree, containing $L-1$ elements. 

The first $L+1$-element cluster $C$ can be obtained as the union of two neighboring $L$-element clusters $A$ and $B$ of $G_L$, which we will call prior subsets. These neighboring clusters $A$ and $B$ are defined by the one-step neighborhood matrix. A new vertex $v$ is only added to the $(L+1)$-order cherry tree we are constructing if a $w$ neighbor of $v$ is already in the $(L+1)$-order cherry tree. These $w$ vertices are listed in the F list of the algorithm. This means that there exists an $L+1$ element cluster $C = A \cup B$ which contains two prior $L$-element subsets $A$ and $B$ such that one of them contains a vertex which is $1$ neighbor apart from the new vertex $v$ in the one-step neighborhood matrix. By joining the new vertex $v$ to a prior subset ($A$ or $B$) we obtain a new $L+1$-element cluster, which we then add to the existing $(L+1)$-order cherry tree. This new cluster will contain two prior subsets which the algorithm joined.

By this procedure, each cluster will be connected to the rest of the tree by at most two prior subsets of $G_L$, which act as separators of $G_{L+1}$. This completes the proof. $\square$
\end{pf}

\begin{figure}[h!]
	\centering
	%\vspace{-1cm}
    \includegraphics[width=0.85\textwidth]{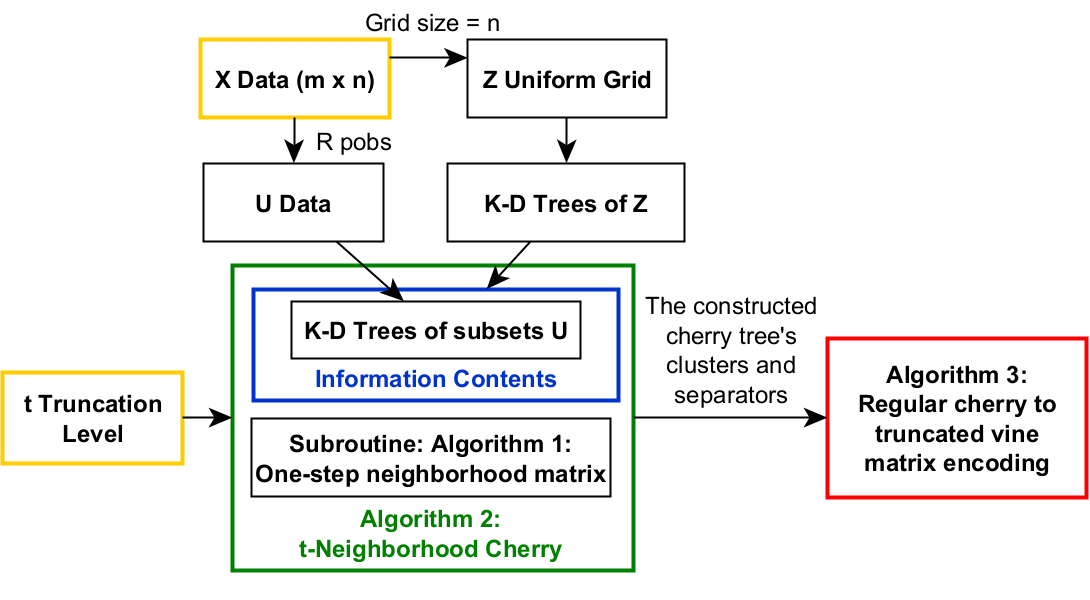}
	\caption{The flowchart of the Trunc-Opt approach. Yellow denotes the inputs, red the denotes the output.}
	\label{flowchart}
\end{figure}

\subsection{Complexities of the algorithms}

\subsubsection{Complexity of calculating the one-step neighborhood matrix}

For a given graph $G=(V,E)$ with $|V|=n$, finding the neighborhood ($\text{Ne}$) of a given vertex takes $O(n)$ time. Calculating $\text{Ne}(s) \cap \text{Ne}(t) \setminus \{s\} \setminus \{t\}$ takes $O(2n)+O(n\log(n))=O(n\log(n))$ time. The intersection of two sets can be calculated via sorting the sets, which has $O(n\log(n))$ time complexity.

The algorithm does the above $O(n\log(n))$ step for each pair of $O(n^2)$ vertices, therefore it is an $O(n^3\log(n))$ algorithm.

\subsubsection{Complexity of the $I$ information content function}

% GYŐRFI LÁSZLÓ: INFORMÁCIÓ TARTALOM KONZISZTENCIA CIKK ????????????????????????????????????????????????????????????????????????????????????

Let $d \in \{1,\dots,n\}$ be given. To calculate $I(U_1\dots,U_d)\approx\widehat{\text{KL}}(U,Z)$, we need to find the $k$'th nearest neighbor of each sample $U^i$, to the rest of the samples $\{U^j\}_{j\ne i}$ (denoted by $\rho_k(i)$), and the $k$'th nearest neighbor to the points of the grid $Z$ (denoted by $\nu_k(i)$).

The average time complexity of setting up a K-D Tree of $m$ vertices in $d$ dimensions is $O(dm\log(dm))$ (see \cite{bentley1975multidimensional}). We are pre-calculating the K-D Trees of all grids of size $2$ to $t$, which will take a total time of

$$\sum_{d=2}^t O(dm\log(dm)) = O\left(m\log(m)\sum_{d=2}^t d + m\sum_{d=2}^t d\log(d)\right) = O(m\log(m)t^2+mt^2\log(t)) = O(mt^2\log(mt))$$

Then to calculate a single Information Content, all grid K-D Trees will be given, afterwards which we will still need to set up the K-D Trees for the $U=(U_1,\dots,U_d)$ dataset. Similarly to before, this will take $O(dm\log(dm))$ time.

Then calculating $k$'th nearest neighbor in either K-D Tree takes $O(\log(dm))$ time (see \cite{bentley1975multidimensional}).

In Formula (\ref{eq:infKL}) for approximating the Information Content, we need to do the above $O(\log(dm))$ calculation $2m$ times, then divide the result, take the base-$2$ logarithm of it, add up each value, and multiply it by $d/m$, all of which take at most an additional $O(m)$ time.

Therefore, estimating $I(U_1,\dots,U_d)$ takes $O(dm\log(dm)+m\log(dm)+m)=O(dm\log(dm))$ steps. The complexity of setting up the additional K-D Trees for the grids will be added to the final algorithm's complexity.

\subsubsection{Complexity of the $t$-neighborhood cherry algorithm}

In Row $1$ of the algorithm we calculate the matrix of Mutual Informations. This matrix has $n \times n$ elements, each taking $O(dm\log(dm))=O(2m\log(2m))=O(m\log(m))$ steps to calculate. Overall, calculating this matrix takes $O(n^2m\log(m))$ steps.

Rows $2-3$ calculate the maximum weight spanning tree of $G_2$. This can be done with Kruskal's algorithm, whose runtime is $O(|E|\log(|E|))$, where $|E|$ is the number of edges of our complete graph, which is on the order of $O(n^2)$. So the total runtime of rows $2-3$ is $O(n^2 \log(n^2)) = O(n^2 \log(n))$.

Row $4$ sums up the calculated Mutual Informations for each edge of the $G_2$ spanning tree in $O(n)$ steps, and row $5$ takes $O(1)$ step.

The rest of the algorithm builds up the cherry tree sequence from level $3$ to $t$. We will be focusing on a specific level, $L$.

Rows $6-11$ take $O(1)$ step. Row $12$ calculates the one-step neighborhood matrix, which we have seen takes $O(n^3\log(n))$ steps.

Rows $13-35$ build the cherry tree for a specific level, $L$, by adding $1$ cluster at a time. We will now be focusing on the complexity of adding $1$ cluster on level $L$. At the end, we will sum the complexity for all clusters added.

Rows $13-19$ take $O(1)$ step. Rows $20-24$ require going through the $A$ one-step neighborhood matrix in $O(n^2)$ steps. In each step, we are adding (at most) $L$ elements to $P$, so the complexity of these rows is $O(n^2L)$.

Rows $25-27$ are used in the first step, otherwise, from $r=2$ to $n-t+1$, rows $28-31$ are used. These steps calculate Information Contents of size $L$ and $L-1$ (except for the first step, where only size $L$ are calculated). An Information Content of size $L$ takes $O(Lm\log(Lm))$ steps to calculate. We are doing this calculation $|P\setminus\mathcal{K}_L|$ times.

$P$ is the set of potential clusters. They are found by the existing elements of the one-step neighborhood matrix. First we consider all possible cliques of size $L$ that can be formed by adding a single edge to the existing graph. Thereafter, we will only consider cliques that connect to the already found indices, represented by set $F$. In the first step, there may be up to $O(n^2)$ potential clusters, since that is how many edges may be missing from $G_L$. Thereafter, we may only consider edges that join to the found vertices, $F$, whose size in step $r$ is $r+2$. There are at most $(r+2)(n-r-3)=O(rn)$ such edges.

So the overall complexity of rows $25-27$ (with $r=1$) is $O(n^2Lm\log(Lm))$, while for rows $28-31$ (with $r\in\{2,\dots,n-t+1\})$ is $O(rn(L+L+1)m\log((L+L+1)m))=O(rnLm\log(Lm))$.

Then rows $32-35$ can be done in at most $O(n^2)$ steps as well, as at most all indices from $1$ to $n$ need to be compared to all elements of $F$.

Exiting the inner for-loop, row $36$ takes at most $2n^2$ steps, as that is at most how many indices $E$ contains, which is an $O(n^2)$ amount.

So the total complexity of the $k$-neighborhood cherry algorithm is

\begin{align*}
&O(n^2m\log(m))+O(n^2\log(n))+O(n)+\sum_{L=3}^t \biggl( O(n^3\log(n)) + O(n^2L+n^2Lm\log(Lm)) + O(n^2) + \\
&+\sum_{r=2}^{n-t+1}O(n^2L)+O(rnLm\log(Lm)) + O(n^2) \biggr) = O(n^2m\log(m)) + \sum_{L=3}^t \biggl( O(n^3\log(n)+n^2Lm\log(Lm)) + \\
&+O((n-t)n^2L) +O((n-t)^2nLm\log(Lm))+O((n-t)n^2)\biggr) = O(n^2m\log(m)) + \\
&+\sum_{L=3}^t \biggl( O(n^3\log(n) + Lm\log(Lm)(n^2+(n-t)^2n) + (n-t)n^2L )\biggr) = O(n^2m\log(m)) + \\
&+O(tn^3\log(n)+t^2m\log(tm)(n^2+(n-t)^2n)+t^2(n-t)n^2) = O(t^2n^3\log(n)+t^2m\log(tm)(n^2+(n-t)^2n))
\end{align*}

The complexity of calculating the K-D Trees of all grids is $O(mt^2\log(mt))$, which is less than the second term of the above time complexity, so adding it will not change this amount. We can further simplify the result by noticing that $n^3 > n^2+(n-t)^2n$. So the final time complexity of the $t$-neighborhood cherry algorithm is

$$O(n^3t^2m\log(tm))$$

\subsubsection{Complexity of the cherry to vine matrix encoding algorithm}

Rows $1$ and $3$ take $O(1)$ steps, while row $2$ finds a leaf node in $\mathcal{K}_t$. This cluster set contains $O(t)$ clusters, each of which contains $t$ elements, so this step take at most $O(t^2)$ comparisons.

Rows $4-7$ finds the initial set of clusters to append to the bottom right of the output matrix. This takes $t-1$ steps, and in each step, we need to find a cluster $K \in \mathcal{K}_i$ so that all of its elements are in $P$. In step $i \in \{2,\dots,t\}$, the cluster $K \in \mathcal{K}_i$ contains $i$ elements. Checking if they all exist in $P$ takes $i$ comparisons. Doing the subtraction $K \setminus P$ then takes $i$ additional comparisons, as $P$ also has $i$ elements at this point. Overall, the complexity of Rows $4-7$ is $\sum_{i=2}^t 2i = 2\left(\frac{t(t+1)}{2}-1\right) = O(t^2)$.

Rows $8-10$ then attach the found elements to the output matrix, which takes $O(t)$ steps.

Rows $11-12$ are $O(1)$ step each.

Rows $13-28$ fill out the lower right triangle of the output matrix, namely columns $n-t+1$ to $n$. The lower triangular matrix in this area contains $O(t^2/2)=O(t^2)$ elements to fill out that the $i$ and $j$ indices go through. For each of these elements, rows $16-26$ is calculated. In the first section (rows $16-21$), we find a cluster in $\mathcal{K}_j$ that connects to our previously found cluster. $K \in \mathcal{K}_j$ and $S$ both contain $O(j)$ elements, so this for-loop takes $\sum_{K \in \mathcal{K}_j}O(j)$ steps. Since $\mathcal{K}_j$ contains $O(j)$ elements, rows $16-21$ take $O(j^2)$ steps. Then, rows $22-23$ are $O(1)$, and rows $24-26$ take $i$ subtractions of size $O(j)$ sets, which is an $O(ij)$ amount of steps. Overall, the complexity of rows $13-28$ is
$$\sum_{i=1}^{t-2} \sum_{j=2}^{i+2} O(j^2)+O(ij) = O\left(\sum_{i=1}^t \sum_{j=1}^i j^2 + \sum_{i=1}^t \sum_{j=1}^i ij\right) = O\left(\sum_{i=1}^t i^3+\sum_{i=1}^t i^3\right)=O(t^4)$$

Rows $29-50$ fill out the rest of the matrix, up to the truncation level. First, for each column, we find the diagonal element in rows $30-36$. For each column, this takes $t$ subtractions of sets of size $O(t)$ and $O(t+i)$, which is overall $O(t(t+i))$ steps. Then in rows $37-44$, we find the cluster that joins to the previous cluster. We need to potentially go through every size-$j$ cluster. There are $O(j)$ of them, so this takes $O(j^2)$ comparisons. Finally, rows $45-46$ take $O(1)$ step each, and rows $47-49$ take $O(t)$ subtractions of $O(i)$ size clusters, which is $O(ti)$ steps. Overall, the complexity of rows $29-50$ is
\begin{align*}
&\sum_{i=1}^{n-t} \left( O(t(t+i)) + \sum_{j=2}^t O(j) + O(j^2) + O(ti)\right) = O\left(\sum_{i=1}^{n-t} (O(t^2) + O(ti)) + \sum_{i=1}^{n-t}\sum_{j=1}^t (O(j^2) + O(ti))\right) = \\
&= O\left((n-t)t^2+t(n-t)^2+\sum_{i=1}^{n-t} (t^3 + it^2) \right) = O\left((n-t)t^2+t(n-t)^2+(n-t)t^3+t^2(n-t)^2\right) = \\ &= O((n-t)t^3+t^2(n-t)^2)
\end{align*}
Finally, in rows  $51-53$ we can optionally fill out the rest of the matrix with  Algorithm $2$ of \cite{pfeifer2024vine}. This section of the matrix contains $n-t$ rows and $n-t$ columns. The complexity of the linked algorithm can be found on Page 18, Theorem 18. With our notations, $n$ is $n-t$, and since in this case, we are not truncating, $t$ also becomes $n-t$, so the complexity of this section is $O((n-t)^3)$.

So the total complexity of of the Cherry to vine matrix encoding algorithm is

\begin{align*}&O(t^2)+O(t^2)+O(t^4)+O((n-t)t^3+t^2(n-t)^2)+O((n-t)^3) = \\
&= O(t^4 +t^3(n-t)+t^2(n-t)^2 + (n-t)^3) \stackrel{(*)}{=} O((n-t)^3)
\end{align*}

Where $(*)$ holds if $O(t^2) \le O(n-t)$, which is usually true, since the truncation level $t$ is usually much smaller than $n$.

\section{Application and comparison to existing vine copula fitting methods}\label{section:application}

\subsection{Vine building algorithms}
Two greedy vine copula construction strategies have been proposed in the literature: a "bottom-up" method by \cite{kurowicka2010optimal} and a "top-down" approach by \cite{Dissmann2013selecting}. Both strategies proceed sequentially, tree by tree and respect the proximity condition (in our terminology, regularity condition) in each step. Kurowicka's bottom-up methodology exploits conditional independences between random variables by using partial correlation. According to \cite{kurowicka2010optimal}: 'The “best vine” is the one whose nodes of top trees (tree with the most conditioning) correspond to the smallest absolute values of partial correlations.' We highlight here that she was the first one who considered the idea of exploiting conditional independences when constructing regular vines. Moreover she called the structure obtained this way a truncated vine. However, the method introduced is applicable only for the case of Gaussian copulas.

Dißmann's approach is the most popular one, and uses greedy tree construction at each level. Selecting regular vine trees sequentially "top-down" means that it starts with the selection of the first tree $T_1$ and continues tree by tree up to the last tree $T_{n-1}$. The first tree $T_1$ can be selected as a maximum weight spanning tree. Given that a tree $T_m$, $m \in\{1, ..., n-2\}$, has been selected, the next tree $T_{m+1}$ is chosen to respect the proximity condition. Pair copula families are selected to minimize AIC copula-by-copula and the parameters are set to their individual maximum likelihood estimates (see \cite{czado2019analyzing}). Clearly, this algorithm only makes a locally optimal selection in each step, since the impact on previous and subsequent trees is ignored. The strategy is, however, reasonable for statistical modeling with regular vine copulas. Strong dependences are modeled first.

\cite{gruber2018bayesian} have shown that the greedy Dißmann algorithm in six-dimensional simulation scenarios achieves about at least $75\%$ of the true log-likelihood, which indicates a good performance, even having the huge number of possible regular vines in mind. \cite{brechmann2012truncated} published the truncated vine constructing algorithm, which was later improved in \cite{brechmann2015truncation}. Other, nonparametric vine copula estimation options and their comparisons can be found in \cite{nagler2017nonparametric}.

% példák, alkalmazás

\subsection{Datasets}
\label{subsection:data}

Here we will list the short description of the datasets we have used in this paper. We are discarding all discrete attributes, and showing the remaining number of attributes in the third column.

\bgroup
\def\arraystretch{0.75}
\begin{center}
%\centering
\begin{tabular}{ |c|c|c|c|c| }
 \hline
 & & Num. of & Num. of \\ 
 Dataset & Discarded & attributes & entries  \\ 
 & columns & (columns) & (rows) \\ 
 \hline &&&\\[-2mm]
 \cite{telescope_link}  & class & 10 & 19020 \\
 \cite{redwine_link} & quality & 11 & 1599 \\  
 \cite{abalone_link} & rings, sex & 7 & 4177 \\
 \cite{wdbc_link} & id, diagnosis & 30 & 569 \\
 \hline
\end{tabular}
\end{center}

\newpage
\subsection{Application of the Trunc-Opt method and comparison to Brechmann's method}

For all four datasets, we have used the following methodology to compare the TruncOpt methodology to Brechmann's truncated vine methodology:

\begin{enumerate}
    \item For all truncation levels $t=3$ to $\min(n,20)$, where $n$ is the number of variables (columns) of the input dataset, we ran Steps $2-6$. (For truncation levels $t > 20$, due to the size of grid generated, the program became infeasible to run on $4$ GB RAM.)
    \item We have applied the TruncOpt methodology (detailed in Section \ref{section:greedyalg}), obtaining the regular cherry to truncated vine matrix $A$, as well as all clusters $\mathcal{C}$ and separators $\mathcal{S}$ the cherry tree sequence contains.
    \item For the clusters on the highest level $t$ ($C_{t,1},\dots,C_{t,n-t+1}$), and separators on the highest level $t$ ($S_{t,1},\dots,S_{t,n-t}$), we have calculated $w(T)=\sum_{i=1}^{n-t+1} I(\mathbf{U}_{C_{t,i}})-\sum_{i=1}^{n-t} I(\mathbf{U}_{S_{t,i}})$, which is the total weight of the $t$'th tree $T$.
    \item We have used Brechmann's method on the input dataset (the RVineStructureSelect function in the R VineCopula package, see \cite{rvine})  with the parameters
    \begin{itemize}
        \item trunclevel $= t-1$ (this returns a vine truncated to level $t$)
        \item selectioncrit $=$ logLik (or log-Likelihood)
        \item progress $=$ TRUE
        \item treecrit $=$ tau
        \item presel $=$ FALSE
        \item cores $= 2$
    \end{itemize}
    to obtain the truncated vine matrix $B$, and the goodness of fit with the log-Likelihood measure.
    \item We have used Brechmann's method on the input dataset and matrix $A$ (the RVineCopSelect function in the R VineCopula package, see \cite{rvine}) with the same parameters as before, and with the additional Matrix $=A$ parameter to obtain the goodness of fit with the log-Likelihood measure, for the TruncOpt case.
    \item Finally, we have read the clusters and separators we obtained with Brechmann's method from matrix $B$ and calculated the weight of the cherry tree generated by these clusters.
\end{enumerate}

Overall, we have obtained the Weight of the truncated vine and log-Likelihood of both the TruncOpt and Brechmann methods, for all datasets, on all possible truncation levels (up to $20$).

\begin{figure}[h!]
	\centering
	%\vspace{-1cm}
	\includegraphics[width=1.0\textwidth]{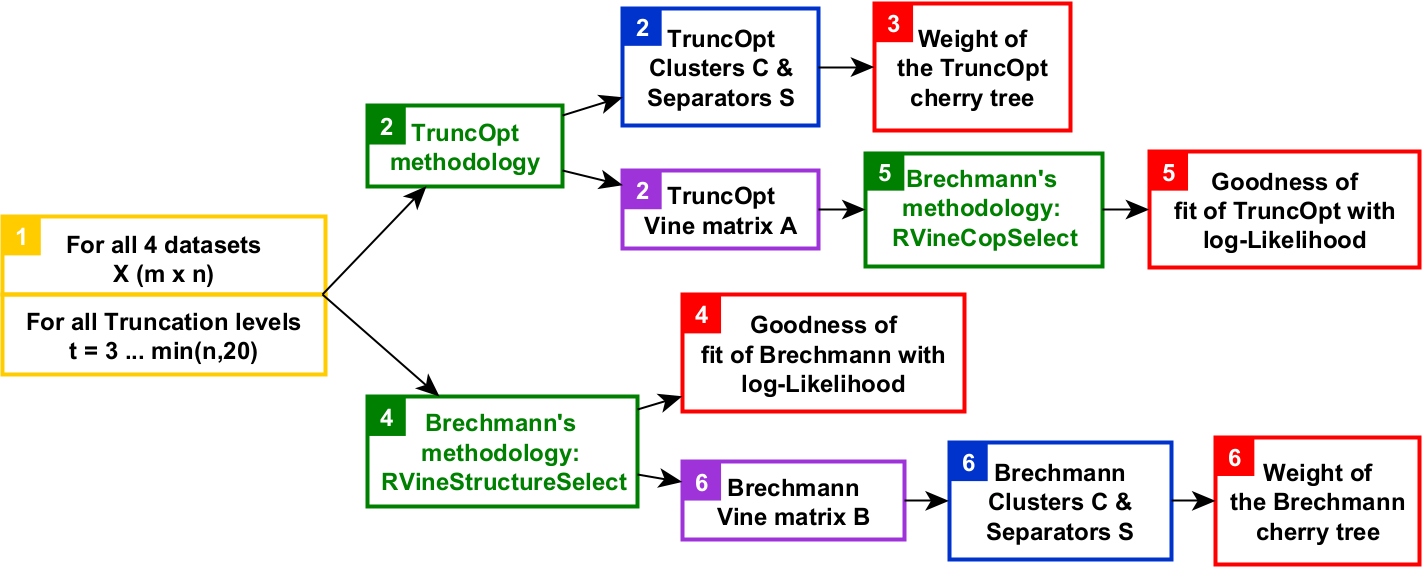}
	\caption{The flowchart of the Trunc-Opt to Brechmann comparison. The numbers in the top left corner of each box represent the previously mentioned comparison steps. Yellow denotes the inputs, red the denotes the outputs - the four measures calculated. Green denotes the algorithms used.}
	\label{test_flowchart}
\end{figure}

\newpage
\subsection{Results}

In this section, we will show $8$ graphs, $2$ for each dataset. Each graph will contain the results of one of the measures (either Weight of the truncated vine or log-Likelihood), and both of the methods (TruncOpt and Brechmann), and show how these measures change as the truncation level $t$ is increasing.

\begin{figure}[h!]
	\centering
	\includegraphics[width=1.0\textwidth]{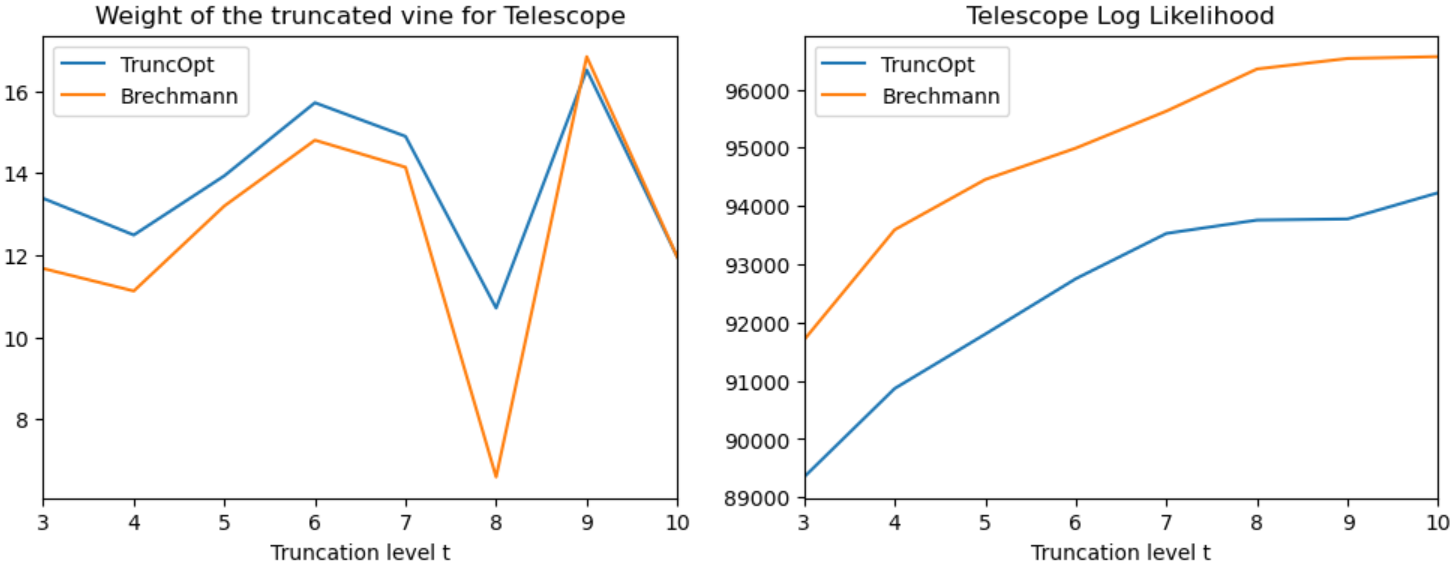}
	\caption{The results for the Telescope dataset. It seems that on most truncation level $t$, the TruncOpt method performs better on Weight of the truncated vine, while Brechmann's method performs better on log-Likelihood. This is no surprise, as these are the measures the two methods are optimizing for.}
	\label{resultsTelescope}
\end{figure}

\begin{figure}[h!]
	\centering
	\includegraphics[width=1.0\textwidth]{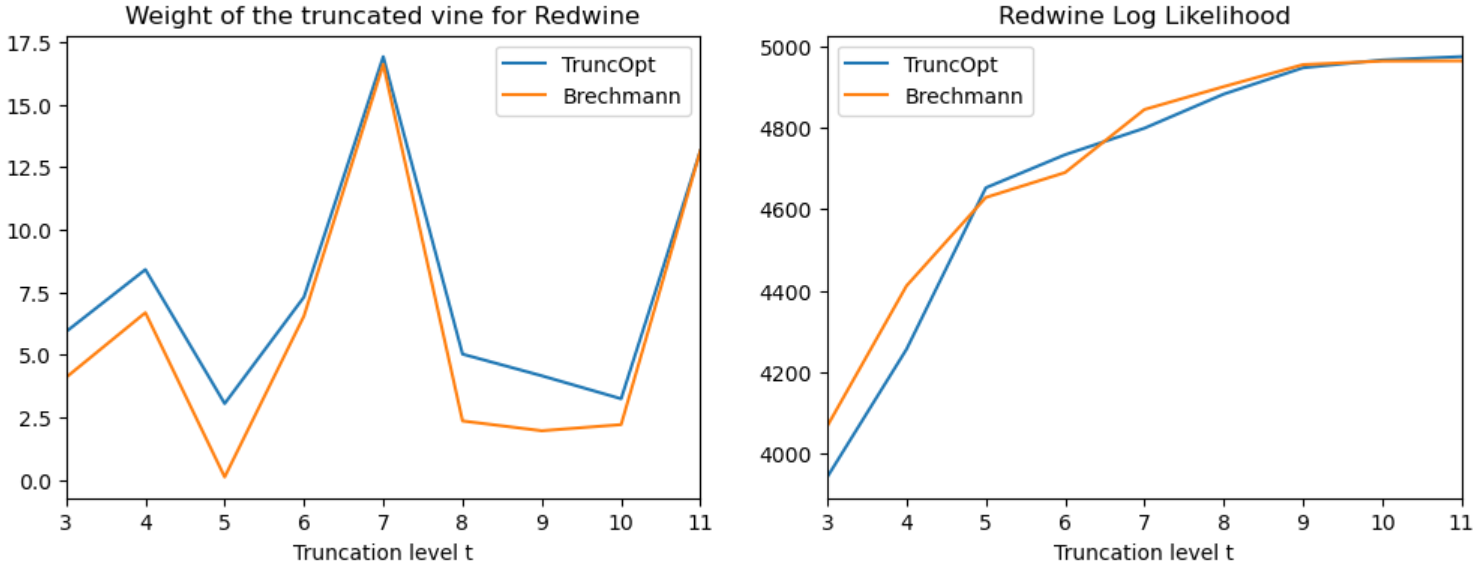}
	\caption{The results for the Redwine dataset. On every truncation level $t$, the TruncOpt method performs better on Weight of the truncated vine, while on log-Likelihood, both methods perform about equally.}
	\label{resultsRedwine}
\end{figure}

\begin{figure}[h!]
	\centering
	\includegraphics[width=1.0\textwidth]{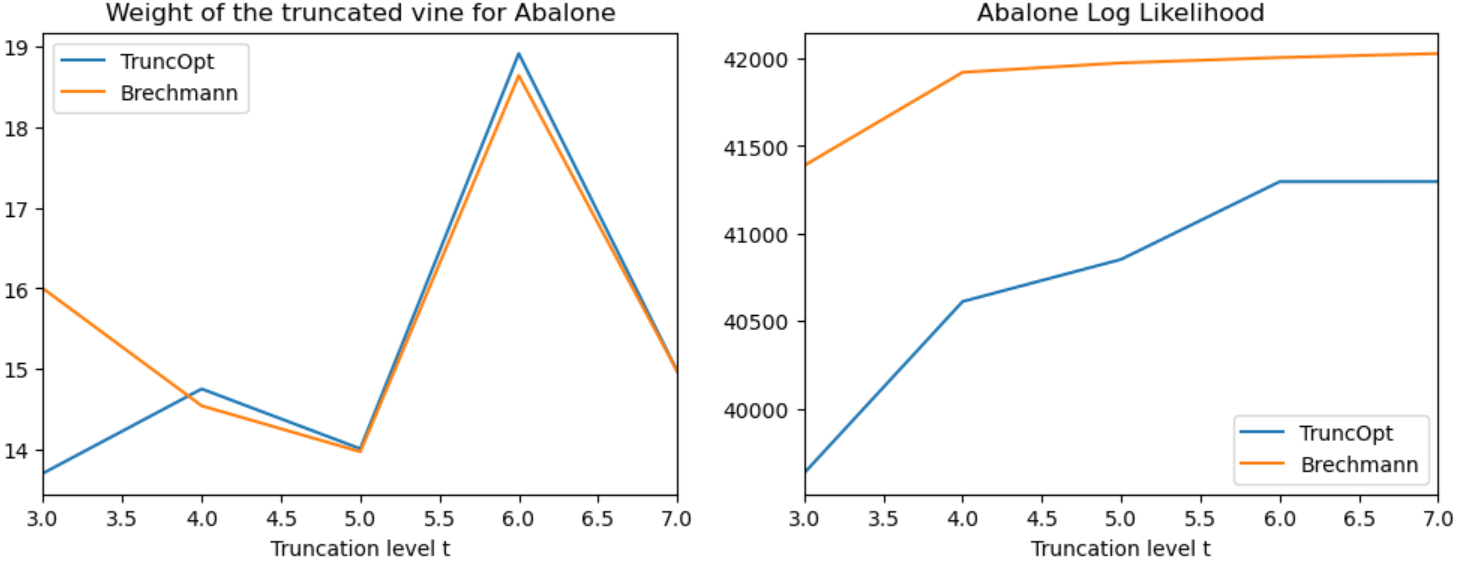}
	\caption{The results for the Abalone dataset. Interestingly, Brechmann's method outperforms the TruncOpt method on level $t=3$ on Weight of the truncated vine, after which the TruncOpt method performs slightly better. However, on log-Likelihood, Brechmann's method consistently performs better.}
	\label{resultsAbalone}
\end{figure}

\begin{figure}[h!]
	\centering
	\includegraphics[width=1.0\textwidth]{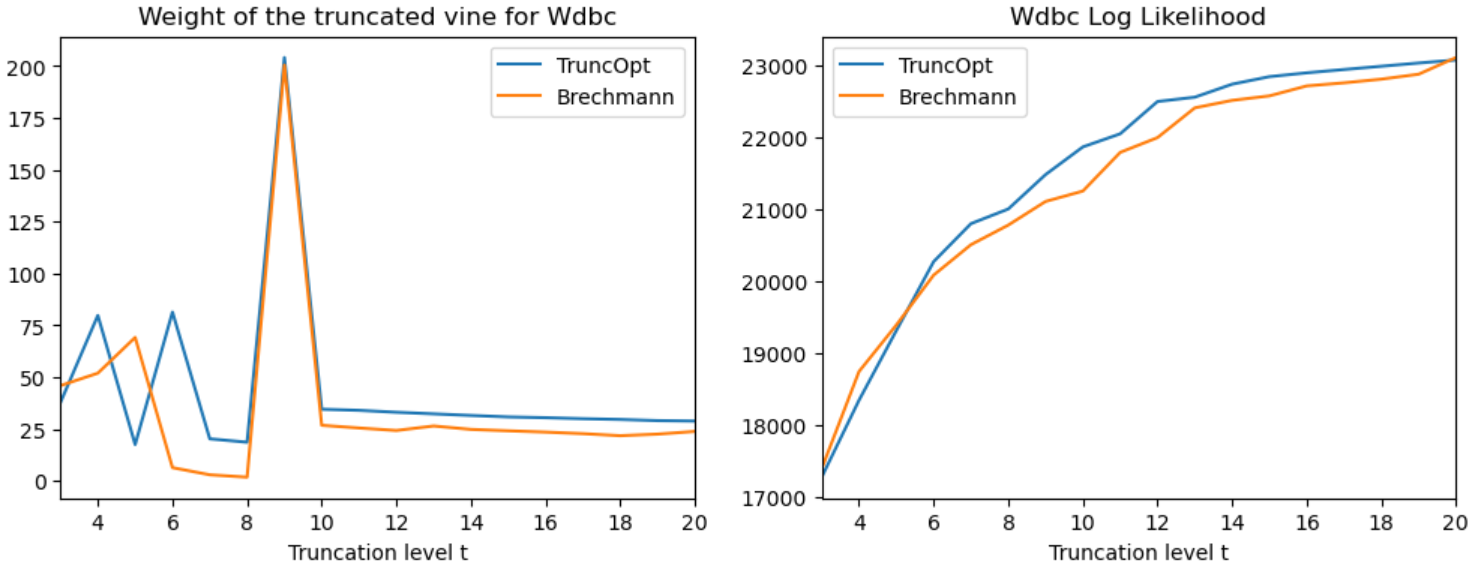}
	\caption{The results for the Wdbc dataset. The TruncOpt method performs better on every truncation level $t$ on Weight of the truncated vine other than $t=3$ and $t=5$, while for truncation levels after $t \ge 6$ on log-Likehood. Interestingly, on truncation level $t=9$, both methods produced an extraordinarily high truncated vine weight.}
	\label{resultsWdbc}
\end{figure}

In Appendix Section \ref{appendix:comp}, we have listed the exact values plotted on these graphs, and the proportions between the TruncOpt and Brechmann methods, for Weight of the truncated vine and log-Likelihood.

\newpage
\section{Conclusions}\label{section:conclusions}

% Edith + Dani
% (nagyon rövid), köszönetnyilvánítás

The current paper suggests a new method for building truncated vine copulas with the novel idea of using conditional independences. We have introduced a method for calculating information content for continuous random variables. We have proved that for a given regular cherry tree there exists a truncated vine that describes the same distribution. It follows that the first tree after truncation fully determines the goodness-of-fit, in terms of Kullback-Leibler divergence, between the truncated vine distribution and the distribution of the input data. This leads to the notion of weight of the truncated vine as the weight of the associated regular cherry tree after truncation. The weight of a truncated vine, as a scoring method, has a large advantage over the popular log-Likelihood score, as it does not require building the full vine, only the last tree, and it also does not require fitting pair-copulas to the vine in order to determine the structure's goodness-of-fit.

The primary contribution of the paper is the new TruncOpt methodology, which builds a new vine structure by greedily minimizing the Kullback-Leibler divergence between the data and the built structure. We gave a method for the output of this algorithm to be encoded in the popular vine matrix, used by most vine copula packages. We have also calculated the complexities of the introduced algorithms, and compared them to the most well known vine structure building method, Brechmann's algorithm. It seems that in most cases, the TruncOpt methodology performs about as well or better than Brechmann's method.

% vine építés
% feltételes függetlenségek
% truncated vine is cherry
% folytonos information content
% inf content konzisztencia
% TruncOpt algo.
% complexity
% az eredmények mátrixba kódolása
% gyors összehasonlítási módszer az utolsó fa alapján
% összehas. Brechmann-nal

\vspace{5mm}

\noindent {\large\bfseries Acknowledgements}
\vspace{3mm}

The research reported in this paper is part of project no. BME-NVA-02, implemented with the support provided by the Ministry of Innovation and Technology of Hungary from the National Research, Development and Innovation Fund, financed under the TKP2021 funding scheme.

% Use the style of numbering in square brackets.
% If nothing is used, default style will be taken.
%\begin{enumerate}[a)]
%\item 
%\item 
%\item 
%\end{enumerate}  

%\begin{itemize}
%\item 
%\item 
%\item 
%\end{itemize}  

%\begin{description}
%\item[]
%\item[] 
%\item[] 
%\end{description}  

% Figure
%\begin{figure}[<options>]
%	\centering
%		\includegraphics[<options>]{}
%	  \caption{}\label{fig1}
%\end{figure}

%\begin{table}[<options>]
%\caption{}\label{tbl1}
%\begin{tabular*}{\tblwidth}{@{}LL@{}}
%\toprule
%  &  \\ % Table header row
%\midrule
% & \\
% & \\
% & \\
% & \\
%\bottomrule
%\end{tabular*}
%\end{table}

% Uncomment and use as the case may be
%\begin{theorem} 
%\end{theorem}

% Uncomment and use as the case may be
%\begin{lemma} 
%\end{lemma}


%\section{}\label{}

\printcredits

\bibliographystyle{cas-model2-names}

%\bibliography{cas-refs}

\bio{}
\endbio


\appendix\section{Numerical results}\label{appendix:comp}

\vspace{3mm}
\begin{center}
\begin{tabular}{r|c|c|c|c|c|c}
\textbf{Telescope} & \multicolumn{2}{c}{TruncOpt} & \multicolumn{2}{c}{Brechmann} & \multicolumn{2}{c}{Proportion} \\
Trunc. Level              & Weight       & LogLik        & Weight        & LogLik        & Weight Prop.   & LogLik Prop.  \\
\hline &&&&&&\\[-2mm]
3                                      & 13.381       & 89349.25      & 11.67281       & 91703.19      & 1.1463         & 0.9743        \\
4                                      & 12.4885      & 90867.58      & 11.1246       & 93591.42      & 1.0699         & 0.9709        \\
5                                      & 13.9316       & 91800.79      & 13.1967        & 94453.63      & 1.2523         & 0.9719        \\
6                                      & 15.7116      & 92751.66      & 14.7988        & 94992.21      & 1.1906         & 0.9764        \\
7                                      & 14.8916       & 93528.58      & 14.1398        & 95627.54      & 1.0532         & 0.9781        \\
8                                      & 10.71      & 93757.25      & 6.5934       & 96349.82      & 1.6244         & 0.9731        \\
9                                      & 16.5104      & 93776.17      & 16.8338       & 96530.02      & 0.9808         & 0.9715        \\
10                                     & 11.953       & 94226.35      & 11.953       & 96561.64      & 1         & 0.9758       
\end{tabular}

\vspace{3mm}

\begin{tabular}{r|c|c|c|c|c|c}
\textbf{Redwine} & \multicolumn{2}{c}{TruncOpt} & \multicolumn{2}{c}{Brechmann} & \multicolumn{2}{c}{Proportion} \\
Trunc. Level            & Weight       & LogLik        & Weight        & LogLik        & Weight Prop.   & LogLik Prop.  \\
\hline &&&&&&\\[-2mm]
3                & 5.9233       & 3943.367      & 4.0958        & 4068.281      & 1.4462         & 0.9693        \\
4                & 8.4027       & 4256.797      & 6.6808        & 4412.236      & 1.2577         & 0.9648        \\
5                & 3.0506       & 4652.62       & 0.1178        & 4628.822      & 25.8964         & 1.0051        \\
6                & 7.3048       & 4733.736      & 6.5477        & 4689.979      & 1.1156         & 1.0093        \\
7                & 16.9074       & 4798.672      & 16.5998       & 4844.528      & 1.0185         & 0.9905        \\
8                & 5.0255       & 4882.263      & 2.356        & 4900.597      & 2.1331         & 0.9963        \\
9                & 4.1599       & 4946.788      & 1.9647        & 4954.588      & 2.1173         & 0.9984        \\
10               & 3.2425       & 4965.992      & 2.2103        & 4963.312      & 1.467         & 1.0005        \\
11               & 13.1506      & 4974.17       & 13.1506       & 4963.804      & 1         & 1.0021       
\end{tabular}

\vspace{3mm}

\begin{tabular}{r|c|c|c|c|c|c}
\textbf{Abalone} & \multicolumn{2}{c}{TruncOpt} & \multicolumn{2}{c}{Brechmann} & \multicolumn{2}{c}{Proportion} \\
Trunc. Level     & Weight       & LogLik        & Weight        & LogLik        & Weight Prop.   & LogLik Prop.  \\
\hline &&&&&&\\[-2mm]
3                & 13.712       & 39631.06      & 15.9972       & 41387.96      & 0.8572         & 0.9576        \\
4                & 14.7553      & 40611.73      & 14.5461       & 41921.35      & 1.0144         & 0.9688        \\
5                & 14.0149      & 40852.62      & 13.9776       & 41974.95      & 1.0027         & 0.9733        \\
6                & 18.908        & 41296.52      & 18.6359       & 42005.6       & 1.0146         & 0.9831        \\
7                & 14.9733      & 41296.69      & 14.9733       & 42028.54      & 1         & 0.9826       
\end{tabular}

\vspace{3mm}

\begin{tabular}{r|c|c|c|c|c|c}
\textbf{Wdbc} & \multicolumn{2}{c}{TruncOpt} & \multicolumn{2}{c}{Brechmann} & \multicolumn{2}{c}{Proportion} \\
Trunc. Level  & Weight        & LogLik       & Weight        & LogLik        & Weight Prop.   & LogLik Prop.  \\
\hline &&&&&&\\[-2mm]
3             & 37.9186       & 17283.81     & 45.9897       & 17402.32      & 0.8245         & 0.9932        \\
4             & 79.8326       & 18351.94     & 51.9748       & 18750.69      & 1.536         & 0.9787        \\
5             & 17.5468       & 19318.85     & 69.207       & 19396.15      & 0.2535         & 0.996         \\
6             & 81.4486       & 20276.6      & 6.4278       & 20089.45      & 12.6713         & 1.0093        \\
7             & 20.3194       & 20804.01     & 2.983             & 20512.07      & 6.8117              & 1.0142        \\
8             & 18.6933       & 21007.36     & 1.835             & 20783.92      & 10.1871              & 1.0108        \\
9             & 204.1556       & 21485.55     & 200.5122             & 21112.32      & 1.0182              & 1.0177        \\
10            & 34.5827      & 21869.86     & 26.849      & 21256.59      & 1.288         & 1.0289        \\
11            & 34.056       & 22051.57     & 25.5415       & 21793.56      & 1.3334         & 1.0118        \\
12            & 33.1292       & 22500.12     & 24.3384       & 21998.9       & 1.3612         & 1.0228        \\
13            & 32.3626         & 22560.16     & 26.5099       & 22412.2       & 1.2208         & 1.0066        \\
14            & 31.592       & 22740.89     & 24.8648       & 22515.87      & 1.2706         & 1.01          \\
15            & 30.8769       & 22845.38     & 24.1852       & 22578.46      & 1.2767         & 1.0118        \\
16            & 30.5172       & 22898.48     & 23.564       & 22716.68      & 1.2951         & 1.008         \\
17            & 30.0212       & 22944.52     & 22.8109       & 22759.42      & 1.3161          & 1.0081        \\
18            & 29.655       & 22989.81     & 21.8351       & 22810.68      & 1.3581         & 1.0079        \\
19            & 29.1022       & 23034.87     & 22.5838       & 22879.38      & 1.2886         & 1.0068        \\
20            & 28.8954       & 23076.57     & 23.887       & 23111.02      & 1.2097         & 0.9985       
\end{tabular}

\end{center}

\end{document}